\documentclass[preprint,12pt]{elsarticle}
\usepackage{setspace}
\usepackage{mathrsfs}
\usepackage{amsmath}
\usepackage{booktabs}
\usepackage{amssymb}
\usepackage{accents}
\usepackage{statex}
\usepackage{enumerate}
\usepackage{graphicx}
\usepackage{subfigure}
\usepackage{indentfirst}
\usepackage{geometry}
\usepackage{array}
\usepackage{booktabs}
\usepackage{tabularx}
\usepackage{multicol}
\usepackage{multirow}
\usepackage{threeparttable}
\usepackage{tikz,xcolor,hyperref}
\usepackage{ntheorem}
\usepackage{caption}
\hypersetup{
colorlinks=true,
linkcolor=black,
citecolor=black
}
\newtheorem{theo}{Theorem}[section]
\newtheorem{lem}{Lemma}[section]

\newtheorem{defi}{Definition}[section]
\newtheorem{rem}{Remark}[section]

\newtheorem*{pro}{Proof}

\newcommand{\PreserveBackslash}[1]{\let\temp=\\#1\let\\=\temp}

\newcolumntype{C}[1]{>{\PreserveBackslash\centering}p{#1}}
\newcolumntype{R}[1]{>{\PreserveBackslash\raggedleft}p{#1}}
\newcolumntype{L}[1]{>{\PreserveBackslash\raggedright}p{#1}}
\journal{J. Fourier Anal. Appl.}

\begin{document}
\begin{frontmatter}
\title{Standard Heisenberg's uncertainty principles of Cohen's class time-frequency distribution with specific kernels}
\tnotetext[mytitlenote]{This work was supported by the National Natural Science Foundation of China under Grant No 61901223 and the Jiangsu Planned Projects for Postdoctoral Research Funds under Grant No 2021K205B.}
\author[1,2,3]{Zhichao Zhang\corref{cor1}}\ead{zzc910731@163.com}
\cortext[cor1]{Corresponding author; Tel: +86-13376073017.}
\address[1]{School of Mathematics and Statistics, Nanjing University of Information Science and Technology, Nanjing 210044, China}
\address[2]{Center for Applied Mathematics of Jiangsu Province, Nanjing University of Information Science and Technology, Nanjing 210044, China}
\address[3]{Jiangsu International Joint Laboratory on System Modeling and Data Analysis, Nanjing University of Information Science and Technology, Nanjing 210044, China}

\begin{abstract}
Time-frequency concentration and resolution of the Cohen's class time-frequency distribution (CCTFD) has attracted much attention in time-frequency analysis. A variety of uncertainty principles of the CCTFD is therefore derived, including the weak Heisenberg type, the Hardy type, the Nazarov type, and the local type. However, the standard Heisenberg type still remains unresolved. In this study, we address the question of how the standard Heisenberg's uncertainty principle of the CCTFD is affected by fundamental properties. The investigated distribution properties are Parseval's relation and the concise frequency domain definition (i.e., only frequency variables are explicitly found in the tensor product), based on which we confine our attention to the CCTFD with some specific kernels. That is the unit modulus and $\mathbf{v}$-independent time translation, reversal and scaling invariant kernel CCTFD (UMITRSK-CCTFD). We then extend the standard Heisenberg's uncertainty principles of the Wigner distribution to those of the UMITRSK-CCTFD, giving birth to various types of attainable lower bounds on the uncertainty product in the UMITRSK-CCTFD domain. The derived results strengthen the existing weak Heisenberg type and fill gaps in the standard Heisenberg type.
\end{abstract}
\end{frontmatter}
\section{Introduction}\label{sec:1}

Cohen's class time-frequency distribution (CCTFD) \cite{Coh66}, also known as the bilinear time-frequency distribution, is one of the most representative time-frequency distributions in the classical time-frequency analysis \cite{Bog20,Bog19,Gro00}. Thanks to the flexibility of the CCTFD in its kernel selection, it includes particular cases many celebrated traditional time-frequency distributions, such as Wigner distribution (WD) \cite{Wig32}, Margenau-Hill distribution \cite{Mar61}, Kirkwood-Rihaczek distribution (KRD) \cite{Kir33,Rih68}, Born-Jordan distribution \cite{Coh66,Cor18}, Page distribution (PD) \cite{Pag52}, Choi-Williams distribution \cite{Cho89}, spectrogram \cite{Gro00}, and Zhao-Atlas-Marks distribution \cite{Zha90}.

\begin{defi}\label{Def1}
\emph{Let a function $f\in L^2(\mathbb{R}^N)$ and a kernel $\phi(\mathbf{v},\mathbf{y})$. The CCTFD of the function $f$ is defined as
\begin{equation}\label{eq1.1}
\mathrm{C}f(\mathbf{x},\mathbf{w})=\mathcal{F}_{\mathbf{y},2}\left\langle\mathfrak{T}_{\mathcal{P}}\left(f\otimes\overline{f}\right)(\mathbf{z},\mathbf{y}),\overline{\mathcal{F}_{\mathbf{v},1}\phi(\mathbf{x}-\mathbf{z},\mathbf{y})}\right\rangle_{\mathbf{z}}(\mathbf{x},\mathbf{w}),
\end{equation}
where the tensor product $\otimes$, the change of coordinates $\mathfrak{T}_{\mathcal{P}}$, the partial Fourier operator $\mathcal{F}_{\mathbf{v},1}$ with respect to the first variables $\mathbf{v}$, the partial Fourier operator $\mathcal{F}_{\mathbf{y},2}$ with respect to the second variables $\mathbf{y}$, and the inner product $\langle,\rangle_{\mathbf{z}}$ with respect to the variables $\mathbf{z}$ are given by $(f\otimes\overline{g})(\mathbf{z},\mathbf{y}):=f(\mathbf{z})\overline{g(\mathbf{y})}$, $\mathfrak{T}_{\mathcal{P}}h(\mathbf{z},\mathbf{y}):=\sqrt{|\mathrm{det}(\mathcal{P})|}h((\mathbf{z},\mathbf{y})\mathcal{P})=h\left(\mathbf{z}+\frac{\mathbf{y}}{2},\mathbf{z}-\frac{\mathbf{y}}{2}\right)$ with $\mathcal{P}=\begin{pmatrix}\mathbf{I}_N&\mathbf{I}_N\\\frac{\mathbf{I}_N}{2}&-\frac{\mathbf{I}_N}{2}\end{pmatrix}$, $\mathcal{F}_{\mathbf{v},1}h(\mathbf{w},\mathbf{y}):=\left\langle h(\mathbf{v},\mathbf{y}),\mathrm{e}^{2\pi\mathrm{i}\mathbf{v}\mathbf{w}^{\mathrm{T}}}\right\rangle_{\mathbf{v}}$, $\mathcal{F}_{\mathbf{y},2}h(\mathbf{x},\mathbf{w}):=\left\langle h(\mathbf{x},\mathbf{y}),\mathrm{e}^{2\pi\mathrm{i}\mathbf{y}\mathbf{w}^{\mathrm{T}}}\right\rangle_{\mathbf{y}}$, and $\langle\circ,\diamond\rangle_{\mathbf{z}}:=\int_{\mathbb{R}^N}\circ(\mathbf{z})\overline{\diamond^{\mathrm{T}}(\mathbf{z})}\mathrm{d}\mathbf{z}$, respectively. Here, $\mathrm{det}(\cdot)$ denotes the determinant operator for matrices, $\mathbf{I}_N$ denotes the $N\times N$ identity matrix, and the superscripts $\mathrm{T}$ and --- denote the transpose operator and complex conjugate operator, respectively.}
\end{defi}

Essential theories and applications of the CCTFD have been well developed by Cohen, see his milestone review article \cite{Coh89} and textbook \cite{Coh95}. Uncertainty principle of the CCTFD is one of the most important theories in time-frequency analysis \cite{CohSPIE95,Coh01,Gro03}. It is able to characterize the time-frequency concentration \cite{Kor05} and resolution \cite{Eld09} of the CCTFD. The most celebrated result in this field is Flandrin's result \cite{Fla99} which gives a weak Heisenberg's uncertainty principle of the CCTFD, stating that a square integrable function cannot be sharply localized in both the weak time-CCTFD domain and frequency-CCTFD domain. The indispensability of the marginal property of the CCTFD in deducing this result was pointed out by Cohen \cite{Coh95}, indicating that the kernel needs to satisfy $\phi(\mathbf{0},\mathbf{y})=1$ and $\phi(\mathbf{v},\mathbf{0})=1$. Subsequently, Korn proposed Hardy's uncertainty principle of the CCTFD with the marginal property, i.e., a kernel satisfying $\phi(\mathbf{0},\mathbf{y})=1$ and $\phi(\mathbf{v},\mathbf{0})=1$ \cite{Kor05}. He also introduced Nazarov's uncertainty principle of the CCTFD with the energy conservation property, i.e., a kernel satisfying $\phi(\mathbf{0},\mathbf{0})=1$ \cite{Kor05}. Moreover, other types of uncertainty principles of the CCTFD were studied more recently by Boggiatto, Carypis and Oliaro \cite{Bog13,Bog14}. To the best of our knowledge, however, the standard Heisenberg's uncertainty principle of the CCTFD is still an open question. This issue is of great theoretical and practical significance because Heisenberg's version plays a fundamental role in various versions of uncertainty principles and the standard Heisenberg's version strengthens the weak one. Meanwhile, taking into consideration that the existing uncertainty principles of the CCTFD hold only for some specific kernels, the study of the standard Heisenberg's version does not seem to be feasible in a general manner. Namely, we have to focus on the CCTFD with some specific kernels.

Fortunately, the standard Heisenberg's uncertainty principles of the WD are currently derived, (they are not derived directly, but as corollaries of the standard Heisenberg's uncertainty principles of the WDs in metaplectic transform domains, such as the $\tau$-WD \cite{Bog13,Bog14,ZhaSP22}, the kernel function WD \cite{Li14}, the kernel function $\tau$-WD \cite{ZhaSPL22}, the symplectic WD \cite{ZhaSP23}, the $\mathbf{K}$-WD \cite{ZhaTIT23-1}, the matrix WD \cite{Bog20,CorAA20,Gia23}, the free metaplectic WD \cite{ZhaTIT23-2}, the cross metaplectic WD \cite{ZhaTIT23-3}, and the cross $\mathcal{A}$-WD \cite{Cor23-2,Cor23-3,CorSUB,Cor22,Cor23-1,Gia23}), which seem feasible to be extended to those of the CCTFD, because the WD is a generator of the CCTFD. More precisely, the WD is none other than the CCTFD with a unit kernel $\phi(\mathbf{v},\mathbf{y})=1$. The main deducing steps of the standard Heisenberg's uncertainty principles of the WD are reproduced here as follows: Step 1. Define the spread in the time-WD domain and the spread in the frequency-WD domain, and use Parseval's relation to simplify the definitions; Step 2. Use the original time domain definition of the WD to convert the spread in the time-WD domain to the spread in the time domain; Step 3. Use the frequency domain definition of the WD to convert the spread in the frequency-WD domain to the spread in the frequency domain; Step 4. Use the results of Steps 2 and 3 to convert the uncertainty product in the WD domain to the conventional uncertainty product in the Fourier transform (FT) domain; Step 5. Use the classical Heisenberg's uncertainty principles to formulate the standard Heisenberg's uncertainty principles of the WD.

The CCTFD with Parseval's relation, i.e., a kernel satisfying $\left|\phi(\mathbf{v},\mathbf{y})\right|=1$ (see Section~\ref{sec:2}), is crucial to finish Steps 1--3 of the standard Heisenberg's uncertainty principles of the CCTFD. Moreover, the CCTFD with a concise frequency domain definition, of which only frequency variables are explicitly found in the tensor product, i.e., a kernel satisfying $\phi(\mathbf{v},2(\mathbf{z}-\mathbf{t}))=\phi(\mathbf{t})$ (see Section~\ref{sec:3}), is necessary to finish Step 3 of the standard Heisenberg's uncertainty principles of the CCTFD. Therefore, this paper focuses mainly on the CCTFD with a kernel satisfying $\left|\phi\right|=1$ (a unit modulus kernel) and $\phi(\mathbf{v},2(\mathbf{z}-\mathbf{t}))=\phi(\mathbf{t})$ (a $\mathbf{v}$-independent time translation, reversal and scaling invariant kernel).

\begin{defi}\label{Def2}
\emph{The unit modulus and $\mathbf{v}$-independent time translation, reversal and scaling invariant kernel CCTFD (UMITRSK-CCTFD) of the function $f$, denoted by $\mathrm{C}_{\mathrm{UMITRSK}}f$, is the CCTFD of $f$ with a kernel satisfying $\left|\phi\right|=1$ and $\phi(\mathbf{v},2(\mathbf{z}-\mathbf{t}))=\phi(\mathbf{t})$.}
\end{defi}

To sum up, this paper simplifies the open question about the standard Heisenberg's uncertainty principles of the CCTFD to those of the UMITRSK-CCTFD. The main deducing steps of the standard Heisenberg's uncertainty principles of the UMITRSK-CCTFD are similar to those of the WD:
\begin{itemize}

    \item Step 1. Define the spread in the time-UMITRSK-CCTFD domain and the spread in the frequency-UMITRSK-CCTFD domain, and use Parseval's relation to simplify the definitions;

    \item Step 2. Use the original time domain definition of the UMITRSK-CCTFD to convert the spread in the time-UMITRSK-CCTFD domain to the spread in the time domain;

    \item Step 3. Use the frequency domain definition of the UMITRSK-CCTFD to convert the spread in the frequency-UMITRSK-CCTFD domain to a summation of two spreads in the frequency domain;

    \item Step 4. Use the results of Steps 2 and 3 to convert the uncertainty product in the UMITRSK-CCTFD domain to a summation of two uncertainty products in the FT domain;

    \item Step 5. Use the classical Heisenberg's uncertainty principles to formulate the standard Heisenberg's uncertainty principles of the UMITRSK-CCTFD.

\end{itemize}

\indent The main contributions of this paper are summarized as follows:
\begin{itemize}

    \item This paper strengthens the weak Heisenberg's uncertainty principle of the CCTFD with the marginal property and fills gaps in the standard Heisenberg's uncertainty principles of the CCTFD.

    \item This paper points out the indispensability of Parseval's relation and the concise frequency domain definition (i.e., only frequency variables are explicitly found in the tensor product) of the CCTFD for the standard Heisenberg's uncertainty principles, proposing the so-called UMITRSK-CCTFD.

    \item This paper establishes the standard Heisenberg's uncertainty principles of the UMITRSK-CCTFD, including the case of the real-valued function $f$ and the real-valued kernel $\phi=\pm1$, the case of the complex-valued function $f$ and the real-valued or complex-valued kernel $\phi\neq\pm\mathrm{e}^{2\pi\mathrm{i}\varphi_f}$, the case of the real-valued function $f$ and the complex-valued kernel $\phi\neq\pm1$, and the case of the complex-valued function $f$ and the complex-valued kernel $\phi=\pm\mathrm{e}^{2\pi\mathrm{i}\varphi_f}$.

\end{itemize}

The remainder of this paper is structured as follows. In Section~\ref{sec:2}, we revisit Parseval's relation of the CCTFD and introduce the so-called unit modulus kernel CCTFD (UMK-CCTFD). In Section~\ref{sec:3}, we investigate the frequency domain definition of the CCTFD and introduce the so-called $\mathbf{v}$-independent time translation, reversal and scaling invariant kernel CCTFD (ITRSK-CCTFD). In Section~\ref{sec:4}, we explore the uncertainty product in the UMITRSK-CCTFD domain. In Section~\ref{sec:5}, we recall the classical Heisenberg's uncertainty principles, including the case of the real-valued function $f$ and the case of the complex-valued function $f$. In Section~\ref{sec:6}, we deduce the standard Heisenberg's uncertainty principles of the UMITRSK-CCTFD. In Section~\ref{sec:7}, we compare the derived results with the well-known Flandrin's result. In Section~\ref{sec:8}, we draw a conclusion. All the technical proofs of our theoretical results are relegated to the appendix parts.

\section*{List of Abbreviations}
\begin{center}
\begin{tabular}{|l|l|}
\hline
CCTFD & Cohen's class time-frequency distribution \\
\hline
WD & Wigner distribution \\
\hline
KRD & Kirkwood-Rihaczek distribution \\
\hline
PD & Page distribution \\
\hline
FT & Fourier transform \\
\hline
\multirow{2}{*}{UMITRSK-CCTFD} & unit modulus and $\mathbf{v}$-independent time translation, \\
 & reversal and scaling invariant kernel CCTFD \\
\hline
UMK-CCTFD & unit modulus kernel CCTFD \\
\hline
\multirow{2}{*}{ITRSK-CCTFD} & $\mathbf{v}$-independent time translation, \\
 & reversal and scaling invariant kernel CCTFD \\
\hline
\end{tabular}
\end{center}

\section*{List of Symbols}
\begin{center}
\begin{tabular}{|l|l|}
\hline
$\mathcal{F}f$ & FT of $f$ \\
\hline
$\mathrm{C}f$ & CCTFD of $f$ \\
\hline
$\mathrm{C}_{\mathrm{UMITRSK}}f$ & UMITRSK-CCTFD of $f$ \\
\hline
$\lambda_f$ & modulus of $f$ \\
\hline
$\varphi_f$ & phase of $f$ \\
\hline
$L^2(\mathbb{R}^N)$ & class of square integrable functions defined on $\mathbb{R}^N$ \\
\hline
$\phi$ & kernel \\
\hline
$\otimes$ & tensor product \\
\hline
$\mathfrak{T}_{\mathcal{P}}$ & change of coordinates \\
\hline
\end{tabular}
\end{center}

\begin{center}
\begin{tabular}{|l|l|}
\hline
$\mathcal{F}_{\mathbf{v},1}$ & partial Fourier operator with respect to the first variables $\mathbf{v}$ \\
\hline
$\mathcal{F}_{\mathbf{y},2}$ & partial Fourier operator with respect to the second variables $\mathbf{y}$ \\
\hline
$\langle,\rangle_{\mathbf{z}}$ & inner product with respect to the variables $\mathbf{z}$ \\
\hline
$\mathrm{det}(\cdot)$ & determinant operator for matrices \\
\hline
$\mathbf{I}_N$ & $N\times N$ identity matrix \\
\hline
$\mathbf{0}_N$ & $N\times N$ null matrix \\
\hline
$\mathrm{T}$ & transpose operator \\
\hline
--- & complex conjugate operator \\
\hline
$\delta(\cdot)$ & Dirac delta operator for vectors \\
\hline
$\lVert\cdot\rVert_{L^2}$ & $L^2$-norm for functions \\
\hline
$\lVert\cdot\rVert_1$ & $1$-norm for vectors \\
\hline
$\lVert\cdot\rVert_2$ & $2$-norm for vectors \\
\hline
\multirow{3}{*}{$|\cdot|$} & element-wise absolute values for vectors \\
 & absolute value for real numbers \\
 & modulus for complex numbers \\
\hline
$\nabla_{\mathbf{x}}(\cdot)$ & gradient operator for functions defined on $\mathbf{x}\in\mathbb{R}^N$ \\
\hline
$\mathbf{x}_f^0$ & moment vector in the time domain \\
\hline
$x_{f;n}^0$ & $n$th component of $\mathbf{x}_f^0$ \\
\hline
$\mathbf{w}_f^0$ & moment vector in the frequency domain \\
\hline
$\omega_{f;n}^0$ & $n$th component of $\mathbf{w}_f^0$ \\
\hline
$\mathbf{x}_{\mathrm{C}_{\mathrm{UMITRSK}},f}^0$ & moment vector in the time-UMITRSK-CCTFD domain \\
\hline
$\mathbf{w}_{\mathrm{C}_{\mathrm{UMITRSK}},f}^0$ & moment vector in the frequency-UMITRSK-CCTFD domain \\
\hline
$\Delta\mathbf{x}_f^2$ & spread in the time domain \\
\hline
$\Delta\mathbf{w}_f^2$ & spread in the frequency domain \\
\hline
$\Delta\mathbf{x}_{\mathrm{C}_{\mathrm{UMITRSK}},f}^2$ & spread in the time-UMITRSK-CCTFD domain \\
\hline
$\Delta\mathbf{w}_{\mathrm{C}_{\mathrm{UMITRSK}},f}^2$ & spread in the frequency-UMITRSK-CCTFD domain \\
\hline
\end{tabular}
\end{center}

\begin{center}
\begin{tabular}{|l|l|}
\hline
$\Delta\mathbf{x}_f^2\Delta\mathbf{w}_f^2$ & uncertainty product in the FT domain \\
\hline
$\Delta\mathbf{x}_{\mathrm{C}_{\mathrm{UMITRSK}},f}^2\Delta\mathbf{w}_{\mathrm{C}_{\mathrm{UMITRSK}},f}^2$ & uncertainty product in the UMITRSK-CCTFD domain \\
\hline
$\mathrm{Cov}_{\mathbf{x}_f,\mathbf{w}_f}$ & covariance in the time-frequency domain \\
\hline
$\mathrm{COV}_{\mathbf{x}_f,\mathbf{w}_f}$ & absolute covariance in the time-frequency domain \\
\hline
\end{tabular}
\end{center}

\section{Parseval's relation of the CCTFD revisited}\label{sec:2}

\begin{lem}[see \cite{Kor05}, Moyal's relation]\label{Lem1}
\emph{Let $\left|\phi(\mathbf{v},\mathbf{y})\right|=1$. The inner product of the CCTFDs of the functions $f$ and $g$ equals to the square modulus of the inner product of $f$ and $g$. That is
\begin{equation}\label{eq2.1}
\left\langle\mathrm{C}f,\mathrm{C}g\right\rangle_{(\mathbf{x},\mathbf{w})}=\left|\langle f,g\rangle_{\mathbf{x}}\right|^2.
\end{equation}}
\end{lem}

\begin{pro}\label{Pro1}
\emph{See Appendix A.$\hfill\blacksquare$}
\end{pro}

\begin{rem}\label{Rem1}
\emph{Eq.~\eqref{eq2.1} is Moyal formula of the CCTFD, which implies Parseval's relation of the CCTFD:
\begin{equation}\label{eq2.2}
\left\lVert\mathrm{C}f\right\rVert_{L^2}=\left\lVert f\right\rVert_{L^2}^2,
\end{equation}
where $\lVert\cdot\rVert_{L^2}$ denotes the $L^2$-norm for functions, induced by the inner product $\langle,\rangle_{\mathbf{z}}$. It deserves to be underlined that Parseval's relation~\eqref{eq2.2} does not hold for arbitrary kernels. Indeed, it makes senses only if $\left|\phi(\mathbf{v},\mathbf{y})\right|=1$. Since Parseval's relation plays a crucial role in establishing the standard Heisenberg's uncertainty principle, it might be wise to not cast the net too wide and settle for the UMK-CCTFD, i.e., the CCTFD with a kernel satisfying $\left|\phi(\mathbf{v},\mathbf{y})\right|=1$.}
\end{rem}

\begin{defi}\label{Def3}
\emph{The UMK-CCTFD of the function $f$ is the CCTFD of $f$ with a kernel satisfying $\left|\phi(\mathbf{v},\mathbf{y})\right|=1$.}
\end{defi}

The well-known WD, KRD and PD are special cases of the UMK-CCTFD corresponding to $\phi(\mathbf{v},\mathbf{y})=1$, $\phi(\mathbf{v},\mathbf{y})=\mathrm{e}^{\pi\mathrm{i}\mathbf{v}\mathbf{y}^{\mathrm{T}}}$ and $\phi(\mathbf{v},\mathbf{y})=\mathrm{e}^{2\pi\mathrm{i}\left\lVert\mathbf{y}\right\rVert_1\mathbf{v}}$, respectively. Here, $\lVert\cdot\rVert_1$ denotes the $1$-norm for vectors.

\section{Frequency domain definition of the CCTFD}\label{sec:3}

\begin{lem}\label{Lem2}
\emph{Let $\mathcal{F}f$ be the FT of $f$, $\mathrm{C}f$ be the CCTFD of $f$, and $f\in L^2(\mathbb{R}^N)$. The CCTFD can be rewritten in the frequency domain as
\begin{align}\label{eq3.1}
\mathrm{C}f(\mathbf{x},\mathbf{w})=&2^N\iint_{\mathbb{R}^{N\times N}}\mathrm{e}^{-4\pi\mathrm{i}\mathbf{z}\mathbf{w}^{\mathrm{T}}}\mathcal{F}f(\mathbf{u})\notag\\
&\times\left(\int_{\mathbb{R}^N}\mathcal{F}_{\mathbf{v},1}\phi(\mathbf{x}-\mathbf{z},2(\mathbf{z}-\mathbf{t}))\overline{f(\mathbf{t})}\mathrm{e}^{2\pi\mathrm{i}\mathbf{t}(2\mathbf{w}-\mathbf{u})^{\mathrm{T}}}\mathrm{d}\mathbf{t}\right)\mathrm{e}^{4\pi\mathrm{i}\mathbf{z}\mathbf{u}^{\mathrm{T}}}\mathrm{d}\mathbf{z}\mathrm{d}\mathbf{u}.
\end{align}}
\end{lem}

\begin{pro}\label{Pro2}
\emph{See Appendix B.$\hfill\blacksquare$}
\end{pro}

\begin{rem}\label{Rem2}
\emph{When $\phi(\mathbf{v},2(\mathbf{z}-\mathbf{t}))=\phi(\mathbf{t})$, it follows that $\mathcal{F}_{\mathbf{v},1}\phi(\mathbf{x}-\mathbf{z},2(\mathbf{z}-\mathbf{t}))=\phi(\mathbf{t})\delta(\mathbf{x}-\mathbf{z})$. Thanks to the sifting property of Dirac delta functions, Eq.~\eqref{eq3.1} simplifies to
\begin{equation}\label{eq3.2}
\mathrm{C}f(\mathbf{x},\mathbf{w})=2^{\frac{N}{2}}\mathrm{e}^{-4\pi\mathrm{i}\mathbf{x}\mathbf{w}^{\mathrm{T}}}\mathcal{F}_{\mathbf{u},2}\mathfrak{T}_{\mathcal{Q}}\left(\mathcal{F}f\otimes\overline{\mathcal{F}\left(f\overline{\phi}\right)}\right)(\mathbf{w},-2\mathbf{x}),
\end{equation}
where $\mathcal{F}\left(f\overline{\phi}\right)$ denotes the FT of $f\overline{\phi}$, and $\mathcal{Q}=\begin{pmatrix}\mathbf{0}_N&2\mathbf{I}_N\\\mathbf{I}_N&-\mathbf{I}_N\end{pmatrix}$, (here $\mathbf{0}_N$ denotes the $N\times N$ null matrix). In the original time domain definition of the CCTFD, the tensor product $f\otimes\overline{f}$ is a function of time variables \cite{ZhaTIT23-3,ZhaTIT23-2}, which is crucial to converting the time domain spread found in the standard Heisenberg's uncertainty principle. Dually, in order to convert the frequency domain spread found in the standard Heisenberg's uncertainty principle, it becomes crucial to rewrite the CCTFD to yield a tensor product with only frequency variables \cite{ZhaTIT23-3,ZhaTIT23-2}. Indeed, the CCTFD in the frequency domain, given by \eqref{eq3.2}, works well for this task, because there are only frequency variables explicitly found in the tensor product $\mathcal{F}f\otimes\overline{\mathcal{F}\left(f\overline{\phi}\right)}$. As it is seen, Eq.~\eqref{eq3.2} does not hold for arbitrary kernels, and it makes senses only if $\phi(\mathbf{v},2(\mathbf{z}-\mathbf{t}))=\phi(\mathbf{t})$. We thus confine our attention to the ITRSK-CCTFD, i.e., the CCTFD with a kernel satisfying $\phi(\mathbf{v},2(\mathbf{z}-\mathbf{t}))=\phi(\mathbf{t})$.}
\end{rem}

\begin{defi}\label{Def4}
\emph{The ITRSK-CCTFD of the function $f$ is the CCTFD of $f$ with a kernel satisfying $\phi(\mathbf{v},2(\mathbf{z}-\mathbf{t}))=\phi(\mathbf{t})$.}
\end{defi}

The well-known WD is a special case of the ITRSK-CCTFD corresponding to $\phi=1$.

With Definitions~\ref{Def3} and \ref{Def4}, the main purpose of this paper is to formulate the standard Heisenberg's uncertainty principles of the UMITRSK-CCTFD, i.e., the CCTFD with a kernel satisfying $\left|\phi\right|=1$ and $\phi(\mathbf{v},2(\mathbf{z}-\mathbf{t}))=\phi(\mathbf{t})$, shown in Definition~\ref{Def2}.

\section{Uncertainty product in the UMITRSK-CCTFD domain}\label{sec:4}

The standard Heisenberg's uncertainty principles of the UMITRSK-CCTFD are essentially a family of inequalities that provide the attainable lower bounds for the uncertainty product in the UMITRSK-CCTFD domain. Here, the uncertainty product is the product of spreads in time-UMITRSK-CCTFD and frequency-UMITRSK-CCTFD domains. As a crucial step in establishing the inequalities, this section aims to explore the uncertainty product in the UMITRSK-CCTFD domain.

This section first introduces definitions of the spread in the time-UMITRSK-CCTFD domain and the spread in the frequency-UMITRSK-CCTFD domain. It then converts the spread in the time-UMITRSK-CCTFD domain and the spread in the frequency-UMITRSK-CCTFD domain to the spread in the time domain and a summation of two spreads in the frequency domain, respectively. And on this basis, it converts the uncertainty product in the UMITRSK-CCTFD domain to a summation of two uncertainty products in the FT domain.

\subsection{Definitions of spreads in UMITRSK-CCTFD domains}\label{subsec4.1}

\begin{defi}\label{Def5}
\emph{Let a function $f\in L^2(\mathbb{R}^N)$, and $\mathrm{C}_{\mathrm{UMITRSK}}f$ be the UMITRSK-CCTFD of the function $f$. It is then defined as}

\emph{(i) Spread in the time-UMITRSK-CCTFD domain:
\begin{equation}\label{eq4.1}
\Delta\mathbf{x}_{\mathrm{C}_{\mathrm{UMITRSK}},f}^2=\frac{\left\lVert\left(\mathbf{x}-\mathbf{x}_{\mathrm{C}_{\mathrm{UMITRSK}},f}^0\right)\mathrm{C}_{\mathrm{UMITRSK}}f\right\rVert_{L^2}^2}{\left\lVert\mathrm{C}_{\mathrm{UMITRSK}}f\right\rVert_{L^2}^2},
\end{equation}
where the moment vector in the time-UMITRSK-CCTFD domain: $\mathbf{x}_{\mathrm{C}_{\mathrm{UMITRSK}},f}^0=\frac{\left\langle\mathbf{x}\mathrm{C}_{\mathrm{UMITRSK}}f,\mathrm{C}_{\mathrm{UMITRSK}}f\right\rangle_{(\mathbf{x},\mathbf{w})}}{\left\lVert\mathrm{C}_{\mathrm{UMITRSK}}f\right\rVert_{L^2}^2}$.}

\emph{(ii) Spread in the frequency-UMITRSK-CCTFD domain:
\begin{equation}\label{eq4.2}
\Delta\mathbf{w}_{\mathrm{C}_{\mathrm{UMITRSK}},f}^2=\frac{\left\lVert\left(\mathbf{w}-\mathbf{w}_{\mathrm{C}_{\mathrm{UMITRSK}},f}^0\right)\mathrm{C}_{\mathrm{UMITRSK}}f\right\rVert_{L^2}^2}{\left\lVert\mathrm{C}_{\mathrm{UMITRSK}}f\right\rVert_{L^2}^2},
\end{equation}
where the moment vector in the frequency-UMITRSK-CCTFD domain: $\mathbf{w}_{\mathrm{C}_{\mathrm{UMITRSK}},f}^0=\frac{\left\langle\mathbf{w}\mathrm{C}_{\mathrm{UMITRSK}}f,\mathrm{C}_{\mathrm{UMITRSK}}f\right\rangle_{(\mathbf{x},\mathbf{w})}}{\left\lVert\mathrm{C}_{\mathrm{UMITRSK}}f\right\rVert_{L^2}^2}$.}
\end{defi}

From Parseval's relation \eqref{eq2.2}, there is $\left\lVert\mathrm{C}_{\mathrm{UMITRSK}}f\right\rVert_{L^2}=\left\lVert f\right\rVert_{L^2}^2$, based on which the spread $\Delta\mathbf{x}_{\mathrm{C}_{\mathrm{UMITRSK}},f}^2$ and moment vector $\mathbf{x}_{\mathrm{C}_{\mathrm{UMITRSK}},f}^0$ in the time-UMITRSK-CCTFD domain simplify to \begin{equation}\label{eq4.3}
\frac{\left\lVert\left(\mathbf{x}-\mathbf{x}_{\mathrm{C}_{\mathrm{UMITRSK}},f}^0\right)\mathrm{C}_{\mathrm{UMITRSK}}f\right\rVert_{L^2}^2}{\left\lVert f\right\rVert_{L^2}^4}
\end{equation}
and
\begin{equation}\label{eq4.4}
\frac{\left\langle\mathbf{x}\mathrm{C}_{\mathrm{UMITRSK}}f,\mathrm{C}_{\mathrm{UMITRSK}}f\right\rangle_{(\mathbf{x},\mathbf{w})}}{\left\lVert f\right\rVert_{L^2}^4},
\end{equation}
respectively; the spread $\Delta\mathbf{w}_{\mathrm{C}_{\mathrm{UMITRSK}},f}^2$ and moment vector $\mathbf{w}_{\mathrm{C}_{\mathrm{UMITRSK}},f}^0$ in the frequency-UMITRSK-CCTFD domain simplify to
\begin{equation}\label{eq4.5}
\frac{\left\lVert\left(\mathbf{w}-\mathbf{w}_{\mathrm{C}_{\mathrm{UMITRSK}},f}^0\right)\mathrm{C}_{\mathrm{UMITRSK}}f\right\rVert_{L^2}^2}{\left\lVert f\right\rVert_{L^2}^4}
\end{equation}
and
\begin{equation}\label{eq4.6}
\frac{\left\langle\mathbf{w}\mathrm{C}_{\mathrm{UMITRSK}}f,\mathrm{C}_{\mathrm{UMITRSK}}f\right\rangle_{(\mathbf{x},\mathbf{w})}}{\left\lVert f\right\rVert_{L^2}^4},
\end{equation}
respectively.

\subsection{Spread in the time-UMITRSK-CCTFD domain}\label{subsec4.2}

The original time domain definition of the UMITRSK-CCTFD, given by \eqref{eq1.1}, is a valid tool to convert the spread $\Delta\mathbf{x}_{\mathrm{C}_{\mathrm{UMITRSK}},f}^2$ in the time-UMITRSK-CCTFD domain, because there are only time variables explicitly found in the tensor product $f\otimes\overline{f}$.

\begin{lem}\label{Lem3}
\emph{Let $\mathrm{C}_{\mathrm{UMITRSK}}f$ be the UMITRSK-CCTFD of $f$, and $f,\lVert\mathbf{x}\rVert_2f\in L^2(\mathbb{R}^N)$, (here $\lVert\cdot\rVert_2$ denotes the $2$-norm for vectors). The spread $\Delta\mathbf{x}_{\mathrm{C}_{\mathrm{UMITRSK}},f}^2$ in the time-UMITRSK-CCTFD domain can be converted as
\begin{equation}\label{eq4.7}
\Delta\mathbf{x}_{\mathrm{C}_{\mathrm{UMITRSK}},f}^2=\frac{\Delta\mathbf{x}_f^2}{2},
\end{equation}
where $\Delta\mathbf{x}_f^2=\frac{\left\lVert\left(\mathbf{x}-\mathbf{x}_f^0\right)f\right\rVert_{L^2}^2}{\lVert f\rVert_{L^2}^2}$ ($\mathbf{x}_f^0=\frac{\left\langle\mathbf{x}f,f\right\rangle_{\mathbf{x}}}{\lVert f\rVert_{L^2}^2}$, the moment vector in the time domain) denotes the spread in the time domain, see (i) of Definition~9 in \cite{ZhaTIT23-3}.}
\end{lem}

\begin{pro}\label{Pro3}
\emph{See Appendix C.$\hfill\blacksquare$}
\end{pro}

Eq.~\eqref{eq4.7} indicates that the spread $\Delta\mathbf{x}_{\mathrm{C}_{\mathrm{UMITRSK}},f}^2$ in the time-UMITRSK-CCTFD domain is none other than the spread $\Delta\mathbf{x}_f^2$ in the time domain, regardless of a multiplier $\frac{1}{2}$.

\subsection{Spread in the frequency-UMITRSK-CCTFD domain}\label{subsec4.3}

The frequency domain definition of the UMITRSK-CCTFD, given by \eqref{eq3.2}, is a valid tool to convert the spread $\Delta\mathbf{w}_{\mathrm{C}_{\mathrm{UMITRSK}},f}^2$ in the frequency-UMITRSK-CCTFD domain, because there are only frequency variables explicitly found in the tensor product $\mathcal{F}f\otimes\overline{\mathcal{F}\left(f\overline{\phi}\right)}$.

\begin{lem}\label{Lem4}
\emph{Let $\mathrm{C}_{\mathrm{UMITRSK}}f$ be the UMITRSK-CCTFD of $f$, $\mathcal{F}f$ and $\mathcal{F}\left(f\overline{\phi}\right)$ be the FTs of $f$ and $f\overline{\phi}$, respectively, and $f,\lVert\mathbf{w}\rVert_2\mathcal{F}f,\lVert\mathbf{w}\rVert_2\mathcal{F}\left(f\overline{\phi}\right)\in L^2(\mathbb{R}^N)$. The spread $\Delta\mathbf{w}_{\mathrm{C}_{\mathrm{UMITRSK}},f}^2$ in the frequency-UMITRSK-CCTFD domain can be converted as
\begin{equation}\label{eq4.8}
\Delta\mathbf{w}_{\mathrm{C}_{\mathrm{UMITRSK}},f}^2=\frac{\Delta\mathbf{w}_f^2+\Delta\mathbf{w}_{f\overline{\phi}}^2}{4},
\end{equation}
where $\Delta\mathbf{w}_f^2=\frac{\left\lVert\left(\mathbf{w}-\mathbf{w}_f^0\right)\mathcal{F}f\right\rVert_{L^2}^2}{\lVert f\rVert_{L^2}^2}$ ($\mathbf{w}_f^0=\frac{\langle\mathbf{w}\mathcal{F}f,\mathcal{F}f\rangle_{\mathbf{w}}}{\lVert f\rVert_{L^2}^2}$, the moment vector in the frequency domain) denotes the spread in the frequency domain, (see (ii) of Definition~9 in \cite{ZhaTIT23-3}), and the definition of $\Delta\mathbf{w}_{f\overline{\phi}}^2$ is similar to that of $\Delta\mathbf{w}_f^2$, just by replacing $f$ with $f\overline{\phi}$.}
\end{lem}

\begin{pro}\label{Pro4}
\emph{See Appendix D.$\hfill\blacksquare$}
\end{pro}

Eq.~\eqref{eq4.8} indicates that the spread $\Delta\mathbf{w}_{\mathrm{C}_{\mathrm{UMITRSK}},f}^2$ in the frequency-UMITRSK-CCTFD domain is none other than a summation of the spreads $\Delta\mathbf{w}_f^2,\Delta\mathbf{w}_{f\overline{\phi}}^2$ in the frequency domain, regardless of a multiplier $\frac{1}{4}$.

\subsection{Product of spreads in time-UMITRSK-CCTFD and frequency-UMITRSK-CCTFD domains}\label{subsec4.4}

The uncertainty product in the UMITRSK-CCTFD domain is the product of spreads in time-UMITRSK-CCTFD and frequency-UMITRSK-CCTFD domains, that is, $\Delta\mathbf{x}_{\mathrm{C}_{\mathrm{UMITRSK}},f}^2\Delta\mathbf{w}_{\mathrm{C}_{\mathrm{UMITRSK}},f}^2$. With Lemmas~\ref{Lem3} and \ref{Lem4}, it can be translated to the conventional uncertainty products in the FT domain.

\begin{lem}\label{Lem5}
\emph{Let $\mathrm{C}_{\mathrm{UMITRSK}}f$ be the UMITRSK-CCTFD of $f$, $\mathcal{F}f$ and $\mathcal{F}\left(f\overline{\phi}\right)$ be the FTs of $f$ and $f\overline{\phi}$, respectively, and $f,\lVert\mathbf{x}\rVert_2f,\lVert\mathbf{w}\rVert_2\mathcal{F}f,\lVert\mathbf{w}\rVert_2\mathcal{F}\left(f\overline{\phi}\right)\in L^2(\mathbb{R}^N)$. There is an equality relation between the uncertainty product $\Delta\mathbf{x}_{\mathrm{C}_{\mathrm{UMITRSK}},f}^2\Delta\mathbf{w}_{\mathrm{C}_{\mathrm{UMITRSK}},f}^2$ in the UMITRSK-CCTFD domain and the conventional uncertainty products $\Delta\mathbf{x}_f^2\Delta\mathbf{w}_f^2,\Delta\mathbf{x}_{f\overline{\phi}}^2\Delta\mathbf{w}_{f\overline{\phi}}^2$ in the FT domain:
\begin{equation}\label{eq4.9}
\Delta\mathbf{x}_{\mathrm{C}_{\mathrm{UMITRSK}},f}^2\Delta\mathbf{w}_{\mathrm{C}_{\mathrm{UMITRSK}},f}^2=\frac{\Delta\mathbf{x}_f^2\Delta\mathbf{w}_f^2+\Delta\mathbf{x}_{f\overline{\phi}}^2\Delta\mathbf{w}_{f\overline{\phi}}^2}{8}.
\end{equation}}
\end{lem}

\begin{pro}\label{Pro5}
\emph{Multiplying \eqref{eq4.7} and \eqref{eq4.8} together gives
\begin{equation}\label{eq4.10}
\Delta\mathbf{x}_{\mathrm{C}_{\mathrm{UMITRSK}},f}^2\Delta\mathbf{w}_{\mathrm{C}_{\mathrm{UMITRSK}},f}^2=\frac{\Delta\mathbf{x}_f^2\Delta\mathbf{w}_f^2+\Delta\mathbf{x}_f^2\Delta\mathbf{w}_{f\overline{\phi}}^2}{8}.
\end{equation}
Thanks to $\left|\phi\right|=1$, there is $\Delta\mathbf{x}_f^2=\Delta\mathbf{x}_{f\overline{\phi}}^2$, and then $\Delta\mathbf{x}_f^2\Delta\mathbf{w}_{f\overline{\phi}}^2=\Delta\mathbf{x}_{f\overline{\phi}}^2\Delta\mathbf{w}_{f\overline{\phi}}^2$.$\hfill\blacksquare$}
\end{pro}

Eq.~\eqref{eq4.9} indicates that the product $\Delta\mathbf{x}_{\mathrm{C}_{\mathrm{UMITRSK}},f}^2$ $\Delta\mathbf{w}_{\mathrm{C}_{\mathrm{UMITRSK}},f}^2$ of spreads in time-UMITRSK-CCTFD and frequency-UMITRSK-CCTFD domains is none other than a summation of the products $\Delta\mathbf{x}_f^2\Delta\mathbf{w}_f^2,\Delta\mathbf{x}_{f\overline{\phi}}^2\Delta\mathbf{w}_{f\overline{\phi}}^2$ of spreads in time and frequency domains, regardless of a multiplier $\frac{1}{8}$.

\section{The classical Heisenberg's uncertainty principles}\label{sec:5}

From Lemma~\ref{Lem5}, the standard Heisenberg's uncertainty principles of the UMITRSK-CCTFD are closely related to the classical Heisenberg's uncertainty principles. In the classical setting, the existing lower bounds on the uncertainty product $\Delta\mathbf{x}_f^2\Delta\mathbf{w}_f^2$ in the FT domain are different for the real-valued function $f$ and the complex-valued function $f$. This section first reviews the classical Heisenberg's uncertainty principle for the real-valued function $f$. It then reviews two versions of classical Heisenberg's uncertainty principles for the complex-valued function $f$.

\subsection{Case of the real-valued function $f$}\label{subsec:5.1}

For the real-valued case, the optimal Gaussian function plays a fundamental role in reaching the lower bound. Below is its definition.

\begin{defi}[see \cite{ZhaTIT23-3}, Definition~12 or \cite{ZhaTIT23-2}, Definition~7]\label{Def6}
\emph{A family of optimal Gaussian functions is defined as
\begin{equation}\label{eq5.1}
f(\mathbf{x})=\mathrm{e}^{-\frac{1}{2\zeta}\left\lVert\mathbf{x}-\mathbf{x}_f^0\right\rVert_2^2+\epsilon}
\end{equation}
for $\zeta>0$ and $\epsilon\in\mathbb{R}$.}
\end{defi}

Heisenberg's inequality for the real-valued function on the uncertainty product $\Delta\mathbf{x}_f^2\Delta\mathbf{w}_f^2$ in the FT domain is reproduced here as follows.

\begin{lem}[see \cite{Fol97}, Corollary~2.8]\label{Lem6}
\emph{Let $f$ be a real-valued function, $\mathcal{F}f$ be the FT of $f$, and $f,\lVert\mathbf{x}\rVert_2f,\lVert\mathbf{w}\rVert_2\mathcal{F}f\in L^2(\mathbb{R}^N)$. Assume that $\nabla_{\mathbf{x}}f$ exists at any point $\mathbf{x}\in\mathbb{R}^N$, (here $\nabla_{\mathbf{x}}(\cdot)$ denotes the gradient operator for functions defined on $\mathbf{x}\in\mathbb{R}^N$). There is an inequality with respect to the uncertainty product $\Delta\mathbf{x}_f^2\Delta\mathbf{w}_f^2$ in the FT domain:
\begin{equation}\label{eq5.2}
\Delta\mathbf{x}_f^2\Delta\mathbf{w}_f^2\geq\frac{N^2}{16\pi^2}.
\end{equation}
When $f$ is non-zero almost everywhere, the equality holds if and only if $f$ is the optimal Gaussian function.}
\end{lem}

\begin{rem}\label{Rem3}
\emph{As seen in the right-hand-side of \eqref{eq5.2}, the attainable lower bound $\frac{N^2}{16\pi^2}$ is independent of $f$.}
\end{rem}

\subsection{Case of the complex-valued function $f$}\label{subsec:5.2}

For the complex-valued case, the optimal Gaussian enveloped chirp function plays a fundamental role in reaching the lower bounds. Below is its definition.

\begin{defi}[see \cite{ZhaJFAA23}, Definition~2.4 or \cite{ZhaTIT23-1}, Eq.~(8) or \cite{ZhaSPL21}, Eqs.~(4)--(9)]\label{Def7}
\emph{Let $x_{f;n}^0$ be the $n$th component of the moment vector $\mathbf{x}_f^0$ in the time domain and $\omega_{f;n}^0$ be the $n$th component of the moment vector $\mathbf{w}_f^0$ in the frequency domain. A family of optimal Gaussian enveloped chirp functions is defined as
\begin{equation}\label{eq5.3}
f(\mathbf{x})=\mathrm{e}^{-\frac{1}{2\zeta}\left\lVert\mathbf{x}-\mathbf{x}_f^0\right\rVert_2^2+\epsilon}\mathrm{e}^{2\pi\mathrm{i}\varphi_f(\mathbf{x})}
\end{equation}
with
\begin{equation}\label{eq5.4}
\varphi_f(\mathbf{x})=\frac{1}{2\varepsilon}\sum\limits_{m=1}^N\eta(x_m)\left(x_m-x_{f;m}^0\right)^2+\mathbf{w}_f^0\mathbf{x}^{\mathrm{T}}+\epsilon^{\eta(x_1),\cdots,\eta(x_N)}
\end{equation}
for $\zeta,\varepsilon>0$ and $\epsilon,\epsilon^{\eta(x_1),\cdots,\eta(x_N)}\in\mathbb{R}$, where
\begin{align}\label{eq5.5}
\eta(x_m)=\left\{
\begin{array}{ll}
1,&m\in\mathbf{n}_{j_1}\\
-1,&m\in\mathbf{n}_{j_2}\\
\mathrm{sgn}\left(x_m-x_{f;m}^0\right),&m\in\mathbf{n}_{j_3}\\
-\mathrm{sgn}\left(x_m-x_{f;m}^0\right),&m\in\mathbf{n}_{j_4}
\end{array}
\right.,
\end{align}
and where
\begin{align}\label{eq5.6}
\mathbf{n}_{j_1}=&\left\{n_{11},\cdots,n_{1j_1}\right\}\notag\\
=&\left\{1\leq n\leq N\bigg|\frac{\partial\varphi_f}{\partial x_n}=\frac{1}{\varepsilon}\left(x_n-x_{f;n}^0\right)+\omega_{f;n}^0\right\},
\end{align}
\begin{align}\label{eq5.7}
\mathbf{n}_{j_2}=&\left\{n_{21},\cdots,n_{2j_2}\right\}\notag\\
=&\left\{1\leq n\leq N\bigg|\frac{\partial\varphi_f}{\partial x_n}=-\frac{1}{\varepsilon}\left(x_n-x_{f;n}^0\right)+\omega_{f;n}^0\right\},
\end{align}
\begin{align}\label{eq5.8}
\mathbf{n}_{j_3}=&\left\{n_{31},\cdots,n_{3j_3}\right\}\notag\\
=&\left\{1\leq n\leq N\Bigg|\frac{\partial\varphi_f}{\partial x_n}=\left\{
\begin{array}{ll}
\frac{1}{\varepsilon}\left(x_n-x_{f;n}^0\right)+\omega_{f;n}^0,&x_n\geq x_{f;n}^0\\
-\frac{1}{\varepsilon}\left(x_n-x_{f;n}^0\right)+\omega_{f;n}^0,&x_n<x_{f;n}^0
\end{array}
\right.\right\}
\end{align}
and
\begin{align}\label{eq5.9}
\mathbf{n}_{j_4}=&\left\{n_{41},\cdots,n_{4j_4}\right\}\notag\\
=&\left\{1\leq n\leq N\Bigg|\frac{\partial\varphi_f}{\partial x_n}=\left\{
\begin{array}{ll}
-\frac{1}{\varepsilon}\left(x_n-x_{f;n}^0\right)+\omega_{f;n}^0,&x_n\geq x_{f;n}^0\\
\frac{1}{\varepsilon}\left(x_n-x_{f;n}^0\right)+\omega_{f;n}^0,&x_n<x_{f;n}^0
\end{array}
\right.\right\}
\end{align}
satisfying $\bigcup\limits_{p=1}^{4}\mathbf{n}_{j_p}=\{1,\cdots,N\}$ and $\mathbf{n}_{j_p}\bigcap\mathbf{n}_{j_q}=\emptyset$ for $p\neq q$.}
\end{defi}

The well-known Heisenberg's inequality for the complex-valued function on the uncertainty product $\Delta\mathbf{x}_f^2\Delta\mathbf{w}_f^2$ in the FT domain is reproduced here as follows.

\begin{lem}[see \cite{Coh95}, Eq.~(3.4)]\label{Lem7}
\emph{Let $f=\lambda_f\mathrm{e}^{2\pi\mathrm{i}\varphi_f}$ be a complex-valued function, $\mathcal{F}f$ be the FT of $f$, and $f,\lVert\mathbf{x}\rVert_2f,\lVert\mathbf{w}\rVert_2\mathcal{F}f\in L^2(\mathbb{R}^N)$. Assume that $\nabla_{\mathbf{x}}\lambda_f,\nabla_{\mathbf{x}}\varphi_f$ exist at any point $\mathbf{x}\in\mathbb{R}^N$. There is an inequality with respect to the uncertainty product $\Delta\mathbf{x}_f^2\Delta\mathbf{w}_f^2$ in the FT domain:
\begin{equation}\label{eq5.10}
\Delta\mathbf{x}_f^2\Delta\mathbf{w}_f^2\geq\frac{N^2}{16\pi^2}+\mathrm{Cov}_{\mathbf{x}_f,\mathbf{w}_f}^2,
\end{equation}
where $\mathrm{Cov}_{\mathbf{x}_f,\mathbf{w}_f}=\frac{\left\langle\left(\mathbf{x}-\mathbf{x}_f^{0}\right)f,\left(\nabla_{\mathbf{x}}\varphi_f-\mathbf{w}_f^{0}\right)f\right\rangle_{\mathbf{x}}}{\lVert f\rVert_2^2}$ denotes the covariance in the time-frequency domain, see (i) of Definition~10 in \cite{ZhaTIT23-3}. When $\nabla_{\mathbf{x}}\varphi_f$ is continuous and $\lambda_f$ is non-zero almost everywhere, the equality holds if and only if $f$ is the optimal Gaussian enveloped chirp function with $\mathbf{n}_{j_3}=\mathbf{n}_{j_4}=\emptyset$.}
\end{lem}

The lower bound $\frac{N^2}{16\pi^2}+\mathrm{Cov}_{\mathbf{x}_f,\mathbf{w}_f}^2$ is tighter than $\frac{N^2}{16\pi^2}$, but it is not the tightest one. In our latest work \cite{ZhaSPL21}, we introduced a sharper Heisenberg's inequality for the complex-valued function on the uncertainty product $\Delta\mathbf{x}_f^2\Delta\mathbf{w}_f^2$ in the FT domain through providing by far the tightest lower bound, as shown in the following.

\begin{lem}[see \cite{Dan13}, Eq.~(1.7) or \cite{ZhaSPL21}, Corollary~1]\label{Lem8}
\emph{Let $f=\lambda_f\mathrm{e}^{2\pi\mathrm{i}\varphi_f}$ be a complex-valued function, $\mathcal{F}f$ be the FT of $f$, and $f,\lVert\mathbf{x}\rVert_2f,\lVert\mathbf{w}\rVert_2\mathcal{F}f\in L^2(\mathbb{R}^N)$. Assume that $\nabla_{\mathbf{x}}\lambda_f,\nabla_{\mathbf{x}}\varphi_f$ exist at any point $\mathbf{x}\in\mathbb{R}^N$. There is an inequality with respect to the uncertainty product $\Delta\mathbf{x}_f^2\Delta\mathbf{w}_f^2$ in the FT domain:
\begin{equation}\label{eq5.11}
\Delta\mathbf{x}_f^2\Delta\mathbf{w}_f^2\geq\frac{N^2}{16\pi^2}+\mathrm{COV}_{\mathbf{x}_f,\mathbf{w}_f}^2,
\end{equation}
where $\mathrm{COV}_{\mathbf{x}_f,\mathbf{w}_f}=\frac{\left\langle\left|\mathbf{x}-\mathbf{x}_f^0\right|f,\left|\nabla_{\mathbf{x}}\varphi_f-\mathbf{w}_f^0\right|f\right\rangle_{\mathbf{x}}}{\lVert f\rVert_2^2}$ denotes the absolute covariance in the time-frequency domain, (here the absolute operator $|\cdot|$ applied to vectors denotes the element-wise absolute values), see (iii) of Definition~10 in \cite{ZhaTIT23-3}. When $\nabla_{\mathbf{x}}\varphi_f$ is continuous and $\lambda_f$ is non-zero almost everywhere, the equality holds if and only if $f$ is the optimal Gaussian enveloped chirp function.}
\end{lem}

\begin{rem}\label{Rem4}
\emph{As seen in the right-hand-side of \eqref{eq5.10} and \eqref{eq5.11}, the attainable lower bounds $\frac{N^2}{16\pi^2}+\mathrm{Cov}_{\mathbf{x}_f,\mathbf{w}_f}^2$ and $\frac{N^2}{16\pi^2}+\mathrm{COV}_{\mathbf{x}_f,\mathbf{w}_f}^2$ depend on $f$, and the latter is tighter because of $\mathrm{COV}_{\mathbf{x}_f,\mathbf{w}_f}\geq\left|\mathrm{Cov}_{\mathbf{x}_f,\mathbf{w}_f}\right|$.}
\end{rem}

See Table~\ref{tab1} for a summary of Heisenberg's inequality for the real-valued function $f$ on the uncertainty product in the FT domain and Heisenberg's inequalities for the complex-valued function $f$ on the uncertainty product in the FT domain.
\begin{table}[htbp]
\centering
\caption{\label{tab1}Various types of attainable lower bounds on the uncertainty product in the FT domain}
\begin{tabular}{ccc}
\hline
\multicolumn{1}{c}{\multirow{4}{*}{Uncertainty product}}
&\multicolumn{1}{c}{$f$}
&\multicolumn{1}{c}{Attainable lower bound}\\
\cmidrule(r){2-3}
\multicolumn{1}{c}{\multirow{4}{*}{$\Delta\mathbf{x}_f^2\Delta\mathbf{w}_f^2$}}
&\multicolumn{1}{c}{Real}
&\multicolumn{1}{c}{$\frac{N^2}{16\pi^2}\triangleq\mathfrak{B}^{\mathrm{R}}$}\\
\cmidrule(r){2-3}
&\multicolumn{1}{c}{\multirow{3}{*}{Complex}}
&\multicolumn{1}{c}{$\frac{N^2}{16\pi^2}+\mathrm{Cov}_{\mathbf{x}_f,\mathbf{w}_f}^2\triangleq\mathfrak{B}_f^{\mathrm{C},\mathrm{Cov}}$}\\
\cmidrule(r){3-3}
&
&\multicolumn{1}{c}{$\frac{N^2}{16\pi^2}+\mathrm{COV}_{\mathbf{x}_f,\mathbf{w}_f}^2\triangleq\mathfrak{B}_f^{\mathrm{C},\mathrm{COV}}$}\\
\hline
\end{tabular}
\end{table}

\section{Standard Heisenberg's uncertainty principles of the UMITRSK-CCTFD}\label{sec:6}

By using the classical Heisenberg's uncertainty principles with respect to $f$ and $f\overline{\phi}$, respectively, it is accessible to obtain the standard Heisenberg's uncertainty principles of the UMITRSK-CCTFD. As implied by Table~\ref{tab1}, there are two different kinds of classical Heisenberg's uncertainty principles, one is for the real-valued function, and the other is for the complex-valued function. Therefore, the standard Heisenberg's uncertainty principles of the UMITRSK-CCTFD are divided into four categories:
\begin{itemize}

    \item Case 1. Standard Heisenberg's uncertainty principle of the UMITRSK-CCTFD for the real-valued functions $f,f\overline{\phi}$, implying the real-valued kernel $\phi=\pm1$.

    \item Case 2. Standard Heisenberg's uncertainty principle of the UMITRSK-CCTFD for the complex-valued functions $f=\lambda_f\mathrm{e}^{2\pi\mathrm{i}\varphi_f},f\overline{\phi}=\lambda_{f\overline{\phi}}\mathrm{e}^{2\pi\mathrm{i}\varphi_{f\overline{\phi}}}$, implying the real-valued or complex-valued kernel $\phi\neq\pm\mathrm{e}^{2\pi\mathrm{i}\varphi_f}$.

    \item Case 3. Standard Heisenberg's uncertainty principle of the UMITRSK-CCTFD for the real-valued function $f$ and the complex-valued function $f\overline{\phi}=\lambda_{f\overline{\phi}}\mathrm{e}^{2\pi\mathrm{i}\varphi_{f\overline{\phi}}}$, implying the complex-valued kernel $\phi\neq\pm1$.

    \item Case 4. Standard Heisenberg's uncertainty principle of the UMITRSK-CCTFD for the complex-valued function $f=\lambda_f\mathrm{e}^{2\pi\mathrm{i}\varphi_f}$ and the real-valued function $f\overline{\phi}$, implying the complex-valued kernel $\phi=\pm\mathrm{e}^{2\pi\mathrm{i}\varphi_f}$.

\end{itemize}

\subsection{Case of the real-valued function $f$ and the real-valued kernel $\phi=\pm1$}\label{subsec:6.1}

\begin{theo}\label{The1}
\emph{Let $f$ be a real-valued function, $\mathrm{C}_{\mathrm{UMITRSK}}f$ be the UMITRSK-CCTFD of $f$ with a real-valued kernel $\phi=\pm1$, $\mathcal{F}f$ be the FT of $f$, and $f,\lVert\mathbf{x}\rVert_2f,\lVert\mathbf{w}\rVert_2\mathcal{F}f\in L^2(\mathbb{R}^N)$. Assume that $\nabla_{\mathbf{x}}f$ exists at any point $\mathbf{x}\in\mathbb{R}^N$. There is an inequality with respect to the uncertainty product $\Delta\mathbf{x}_{\mathrm{C}_{\mathrm{UMITRSK}},f}^2\Delta\mathbf{w}_{\mathrm{C}_{\mathrm{UMITRSK}},f}^2$ in the UMITRSK-CCTFD domain:
\begin{equation}\label{eq6.1}
\Delta\mathbf{x}_{\mathrm{C}_{\mathrm{UMITRSK}},f}^2\Delta\mathbf{w}_{\mathrm{C}_{\mathrm{UMITRSK}},f}^2\geq\frac{\mathfrak{B}^{\mathrm{R}}}{4}.
\end{equation}
When $f$ is non-zero almost everywhere, the equality holds if and only if $f$ is the optimal Gaussian function.}
\end{theo}

\begin{pro}\label{Pro6}
\emph{By using Lemma~\ref{Lem6} with respect to the real-valued functions $f,f\overline{\phi}$, it follows that
\begin{equation}\label{eq6.2}
\Delta\mathbf{x}_f^2\Delta\mathbf{w}_f^2=\Delta\mathbf{x}_{f\overline{\phi}}^2\Delta\mathbf{w}_{f\overline{\phi}}^2\geq\mathfrak{B}^{\mathrm{R}}.
\end{equation}
Substituting \eqref{eq6.2} into \eqref{eq4.9} gives the required result \eqref{eq6.1}.$\hfill\blacksquare$}
\end{pro}

\begin{rem}\label{Rem5}
\emph{Theorem~\ref{The1} with $\phi=1$ reduces to Heisenberg's inequality for the real-valued function $f$ on the uncertainty product in the WD domain. The lower bound $\frac{\mathfrak{B}^{\mathrm{R}}}{4}$ is attainable by the optimal Gaussian function.}
\end{rem}

\subsection{Case of the complex-valued function $f$ and the real-valued or complex-valued kernel $\phi\neq\pm\mathrm{e}^{2\pi\mathrm{i}\varphi_f}$}\label{subsec:6.2}

\begin{theo}\label{The2}
\emph{Let $f=\lambda_f\mathrm{e}^{2\pi\mathrm{i}\varphi_f}$ be a complex-valued function, $\mathrm{C}_{\mathrm{UMITRSK}}f$ be the UMITRSK-CCTFD of $f$ with a real-valued or complex-valued kernel $\phi=\mathrm{e}^{2\pi\mathrm{i}\varphi_\phi}\neq\pm\mathrm{e}^{2\pi\mathrm{i}\varphi_f}$, $\mathcal{F}f$ and $\mathcal{F}\left(f\overline{\phi}\right)$ be the FTs of $f$ and $f\overline{\phi}$, respectively, and $f,\lVert\mathbf{x}\rVert_2f,\lVert\mathbf{w}\rVert_2\mathcal{F}f,\lVert\mathbf{w}\rVert_2\mathcal{F}\left(f\overline{\phi}\right)\in L^2(\mathbb{R}^N)$. Assume that $\nabla_{\mathbf{x}}\lambda_f,\nabla_{\mathbf{x}}\varphi_f,\nabla_{\mathbf{x}}\varphi_\phi$ exist at any point $\mathbf{x}\in\mathbb{R}^N$. There is an inequality chain with respect to the uncertainty product $\Delta\mathbf{x}_{\mathrm{C}_{\mathrm{UMITRSK}},f}^2\Delta\mathbf{w}_{\mathrm{C}_{\mathrm{UMITRSK}},f}^2$ in the UMITRSK-CCTFD domain:
\begin{align}\label{eq6.3}
\Delta\mathbf{x}_{\mathrm{C}_{\mathrm{UMITRSK}},f}^2\Delta\mathbf{w}_{\mathrm{C}_{\mathrm{UMITRSK}},f}^2&\geq\frac{\mathfrak{B}_f^{\mathrm{C},\mathrm{COV}}+\mathfrak{B}_{f\overline{\phi}}^{\mathrm{C},\mathrm{COV}}}{8}\notag\\
&\geq\frac{\mathfrak{B}_f^{\mathrm{C},\mathrm{Cov}}+\mathfrak{B}_{f\overline{\phi}}^{\mathrm{C},\mathrm{Cov}}}{8}.
\end{align}
When $\nabla_{\mathbf{x}}\varphi_f,\nabla_{\mathbf{x}}\varphi_{\phi}$ are continuous and $\lambda_f$ is non-zero almost everywhere, the first equality holds if and only if $f$ and $f\overline{\phi}$ are both the optimal Gaussian enveloped chirp functions and the second equality holds if and only if $f$ and $f\overline{\phi}$ are both the optimal Gaussian enveloped chirp functions with $\mathbf{n}_{j_3}=\mathbf{n}_{j_4}=\emptyset$.}
\end{theo}

\begin{pro}\label{Pro7}
\emph{By using Lemma~\ref{Lem7} with respect to the complex-valued functions $f,f\overline{\phi}$, it follows that
\begin{equation}\label{eq6.4}
\Delta\mathbf{x}_f^2\Delta\mathbf{w}_f^2\geq\mathfrak{B}_f^{\mathrm{C},\mathrm{Cov}},\Delta\mathbf{x}_{f\overline{\phi}}^2\Delta\mathbf{w}_{f\overline{\phi}}^2\geq\mathfrak{B}_{f\overline{\phi}}^{\mathrm{C},\mathrm{Cov}}.
\end{equation}
Similarly, with Lemma~\ref{Lem8} there are
\begin{equation}\label{eq6.5}
\Delta\mathbf{x}_f^2\Delta\mathbf{w}_f^2\geq\mathfrak{B}_f^{\mathrm{C},\mathrm{COV}},\Delta\mathbf{x}_{f\overline{\phi}}^2\Delta\mathbf{w}_{f\overline{\phi}}^2\geq\mathfrak{B}_{f\overline{\phi}}^{\mathrm{C},\mathrm{COV}}.
\end{equation}
Substituting \eqref{eq6.4} into \eqref{eq4.9} gives the second inequality in the required result \eqref{eq6.3}. Substituting \eqref{eq6.5} into \eqref{eq4.9} gives the first inequality in the required result \eqref{eq6.3}.$\hfill\blacksquare$}
\end{pro}

\begin{rem}\label{Rem6}
\emph{Theorem~\ref{The2} with $\phi=1$ reduces to Heisenberg's inequalities for the complex-valued function $f$ on the uncertainty product in the WD domain. The lower bound $\frac{\mathfrak{B}_f^{\mathrm{C},\mathrm{COV}}}{4}$ is attainable by the optimal Gaussian enveloped chirp function; the lower bound $\frac{\mathfrak{B}_f^{\mathrm{C},\mathrm{Cov}}}{4}$ is attainable by the optimal Gaussian enveloped chirp function with $\mathbf{n}_{j_3}=\mathbf{n}_{j_4}=\emptyset$.}
\end{rem}

\subsection{Case of the real-valued function $f$ and the complex-valued kernel $\phi\neq\pm1$}\label{subsec:6.3}

\begin{theo}\label{The3}
\emph{Let $f$ be a real-valued function, $\mathrm{C}_{\mathrm{UMITRSK}}f$ be the UMITRSK-CCTFD of $f$ with a complex-valued kernel $\phi=\mathrm{e}^{2\pi\mathrm{i}\varphi_{\phi}}\neq\pm1$, $\mathcal{F}f$ and $\mathcal{F}\left(f\overline{\phi}\right)$ be the FTs of $f$ and $f\overline{\phi}$, respectively, and $f,\lVert\mathbf{x}\rVert_2f,\lVert\mathbf{w}\rVert_2\mathcal{F}f,\lVert\mathbf{w}\rVert_2\mathcal{F}\left(f\overline{\phi}\right)\in L^2(\mathbb{R}^N)$. Assume that $\nabla_{\mathbf{x}}f,\nabla_{\mathbf{x}}\varphi_\phi$ exist at any point $\mathbf{x}\in\mathbb{R}^N$. There is an inequality chain with respect to the uncertainty product $\Delta\mathbf{x}_{\mathrm{C}_{\mathrm{UMITRSK}},f}^2\Delta\mathbf{w}_{\mathrm{C}_{\mathrm{UMITRSK}},f}^2$ in the UMITRSK-CCTFD domain:
\begin{align}\label{eq6.6}
\Delta\mathbf{x}_{\mathrm{C}_{\mathrm{UMITRSK}},f}^2\Delta\mathbf{w}_{\mathrm{C}_{\mathrm{UMITRSK}},f}^2&\geq\frac{\mathfrak{B}^{\mathrm{R}}+\mathfrak{B}_{f\overline{\phi}}^{\mathrm{C},\mathrm{COV}}}{8}\notag\\
&\geq\frac{\mathfrak{B}^{\mathrm{R}}+\mathfrak{B}_{f\overline{\phi}}^{\mathrm{C},\mathrm{Cov}}}{8}.
\end{align}
When $\nabla_{\mathbf{x}}\varphi_{\phi}$ is continuous and $f$ is non-zero almost everywhere, the first equality holds if and only if $f\overline{\phi}$ is the optimal Gaussian enveloped chirp function and the second equality holds if and only if $f\overline{\phi}$ is the optimal Gaussian enveloped chirp function with $\mathbf{n}_{j_3}=\mathbf{n}_{j_4}=\emptyset$.}
\end{theo}

\begin{pro}\label{Pro8}
\emph{By using Lemma~\ref{Lem6} with respect to the real-valued function $f$ and using Lemma~\ref{Lem7} with respect to the complex-valued function $f\overline{\phi}$, it follows that
\begin{equation}\label{eq6.7}
\Delta\mathbf{x}_f^2\Delta\mathbf{w}_f^2\geq\mathfrak{B}^{\mathrm{R}},\Delta\mathbf{x}_{f\overline{\phi}}^2\Delta\mathbf{w}_{f\overline{\phi}}^2\geq\mathfrak{B}_{f\overline{\phi}}^{\mathrm{C},\mathrm{Cov}}.
\end{equation}
Similarly, with Lemmas~\ref{Lem6} and \ref{Lem8} there are
\begin{equation}\label{eq6.8}
\Delta\mathbf{x}_f^2\Delta\mathbf{w}_f^2\geq\mathfrak{B}^{\mathrm{R}},\Delta\mathbf{x}_{f\overline{\phi}}^2\Delta\mathbf{w}_{f\overline{\phi}}^2\geq\mathfrak{B}_{f\overline{\phi}}^{\mathrm{C},\mathrm{COV}}.
\end{equation}
Substituting \eqref{eq6.7} into \eqref{eq4.9} gives the second inequality in the required result \eqref{eq6.6}. Substituting \eqref{eq6.8} into \eqref{eq4.9} gives the first inequality in the required result \eqref{eq6.6}.$\hfill\blacksquare$}
\end{pro}

\subsection{Case of the complex-valued function $f$ and the complex-valued kernel $\phi=\pm\mathrm{e}^{2\pi\mathrm{i}\varphi_f}$}\label{subsec:6.4}

\begin{theo}\label{The4}
\emph{Let $f=\lambda_f\mathrm{e}^{2\pi\mathrm{i}\varphi_f}$ be a complex-valued function, $\mathrm{C}_{\mathrm{UMITRSK}}f$ be the UMITRSK-CCTFD of $f$ with a complex-valued kernel $\phi=\pm\mathrm{e}^{2\pi\mathrm{i}\varphi_f}$, $\mathcal{F}f$ and $\mathcal{F}\left(f\overline{\phi}\right)$ be the FTs of $f$ and $f\overline{\phi}$, respectively, and $f,\lVert\mathbf{x}\rVert_2f,\lVert\mathbf{w}\rVert_2\mathcal{F}f,\lVert\mathbf{w}\rVert_2\mathcal{F}\left(f\overline{\phi}\right)\in L^2(\mathbb{R}^N)$. Assume that $\nabla_{\mathbf{x}}\lambda_f,\nabla_{\mathbf{x}}\varphi_f$ exist at any point $\mathbf{x}\in\mathbb{R}^N$. There is an inequality chain with respect to the uncertainty product $\Delta\mathbf{x}_{\mathrm{C}_{\mathrm{UMITRSK}},f}^2\Delta\mathbf{w}_{\mathrm{C}_{\mathrm{UMITRSK}},f}^2$ in the UMITRSK-CCTFD domain:
\begin{align}\label{eq6.9}
\Delta\mathbf{x}_{\mathrm{C}_{\mathrm{UMITRSK}},f}^2\Delta\mathbf{w}_{\mathrm{C}_{\mathrm{UMITRSK}},f}^2&\geq\frac{\mathfrak{B}_f^{\mathrm{C},\mathrm{COV}}+\mathfrak{B}^{\mathrm{R}}}{8}\notag\\
&\geq\frac{\mathfrak{B}_f^{\mathrm{C},\mathrm{Cov}}+\mathfrak{B}^{\mathrm{R}}}{8}.
\end{align}
When $\nabla_{\mathbf{x}}\varphi_f$ is continuous and $\lambda_f$ is non-zero almost everywhere, the first equality holds if and only if $f$ is optimal Gaussian enveloped chirp function and the second equality holds if and only if $f$ is optimal Gaussian enveloped chirp function with $\mathbf{n}_{j_3}=\mathbf{n}_{j_4}=\emptyset$.}
\end{theo}

\begin{pro}\label{Pro9}
\emph{The proof is similar to that of Theorem~\ref{The3}, and then it is omitted.$\hfill\blacksquare$}
\end{pro}

See Table~\ref{tab2} for a summary of Heisenberg's inequality for the real-valued function $f$ on the uncertainty product in the UMITRSK-CCTFD domain associated with the real-valued kernel $\phi=\pm1$ (i.e., the real-valued function $f\overline{\phi}$), Heisenberg's inequalities for the complex-valued function $f$ on the uncertainty product in the UMITRSK-CCTFD domain associated with the real-valued or complex-valued kernel $\phi\neq\pm\mathrm{e}^{2\pi\mathrm{i}\varphi_f}$ (i.e., the complex-valued function $f\overline{\phi}$), Heisenberg's inequalities for the real-valued function $f$ on the uncertainty product in the UMITRSK-CCTFD domain associated with the complex-valued kernel $\phi\neq\pm1$ (i.e., the complex-valued function $f\overline{\phi}$), and Heisenberg's inequalities for the complex-valued function $f$ on the uncertainty product in the UMITRSK-CCTFD domain associated with the complex-valued kernel $\phi=\pm\mathrm{e}^{2\pi\mathrm{i}\varphi_f}$ (i.e., the real-valued function $f\overline{\phi}$).
\begin{table}[htbp]
\centering
\caption{\label{tab2}Various types of attainable lower bounds on the uncertainty product in the UMITRSK-CCTFD domain}
\scriptsize
\begin{tabular}{ccccc}
\hline
\multicolumn{1}{c}{\multirow{13}{*}{Uncertainty product}}
&\multicolumn{1}{c}{$f$}
&\multicolumn{1}{c}{$\phi$}
&\multicolumn{1}{c}{$f\overline{\phi}$}
&\multicolumn{1}{c}{Attainable lower bound}\\
\cmidrule(r){2-5}
\multicolumn{1}{c}{\multirow{13}{*}{$\Delta\mathbf{x}_{\mathrm{C}_{\mathrm{UMITRSK}},f}^2\Delta\mathbf{w}_{\mathrm{C}_{\mathrm{UMITRSK}},f}^2$}}
&\multicolumn{1}{c}{Real}
&\multicolumn{1}{c}{$\pm1$ (Real)}
&\multicolumn{1}{c}{Real}
&\multicolumn{1}{c}{$\frac{\mathfrak{B}^{\mathrm{R}}}{4}$}\\
\cmidrule(r){2-5}
&\multicolumn{1}{c}{\multirow{2}{*}{Complex}}
&\multicolumn{1}{c}{\multirow{2}{*}{$\neq\pm\mathrm{e}^{2\pi\mathrm{i}\varphi_f}$ (Real or Complex)}}
&\multicolumn{1}{c}{\multirow{2}{*}{Complex}}
&\multicolumn{1}{c}{$\frac{\mathfrak{B}_f^{\mathrm{C},\mathrm{Cov}}+\mathfrak{B}_{f\overline{\phi}}^{\mathrm{C},\mathrm{Cov}}}{8},$}\\
&
&
&
&\multicolumn{1}{c}{$\frac{\mathfrak{B}_f^{\mathrm{C},\mathrm{COV}}+\mathfrak{B}_{f\overline{\phi}}^{\mathrm{C},\mathrm{COV}}}{8}$}\\
\cmidrule(r){2-5}
&\multicolumn{1}{c}{\multirow{2}{*}{Real}}
&\multicolumn{1}{c}{\multirow{2}{*}{$\neq\pm1$ (Complex)}}
&\multicolumn{1}{c}{\multirow{2}{*}{Complex}}
&\multicolumn{1}{c}{$\frac{\mathfrak{B}^{\mathrm{R}}+\mathfrak{B}_{f\overline{\phi}}^{\mathrm{C},\mathrm{Cov}}}{8},$}\\
&
&
&
&\multicolumn{1}{c}{$\frac{\mathfrak{B}^{\mathrm{R}}+\mathfrak{B}_{f\overline{\phi}}^{\mathrm{C},\mathrm{COV}}}{8}$}\\
\cmidrule(r){2-5}
&\multicolumn{1}{c}{\multirow{2}{*}{Complex}}
&\multicolumn{1}{c}{\multirow{2}{*}{$\pm\mathrm{e}^{2\pi\mathrm{i}\varphi_f}$ (Complex)}}
&\multicolumn{1}{c}{\multirow{2}{*}{Real}}
&\multicolumn{1}{c}{$\frac{\mathfrak{B}_f^{\mathrm{C},\mathrm{Cov}}+\mathfrak{B}^{\mathrm{R}}}{8},$}\\
&
&
&
&\multicolumn{1}{c}{$\frac{\mathfrak{B}_f^{\mathrm{C},\mathrm{COV}}+\mathfrak{B}^{\mathrm{R}}}{8}$}\\
\hline
\end{tabular}
\end{table}

\section{Discussion}\label{sec:7}

This section presents a discussion on the main differences and connections between the proposed standard Heisenberg's uncertainty principles and the existing weak one.

The celebrated Flandrin's result is a weak Heisenberg's uncertainty principle of the CCTFD with a kernel satisfying $\phi(\mathbf{0},\mathbf{y})=1$ and $\phi(\mathbf{v},\mathbf{0})=1$. That is
\begin{equation}\label{eq7.1}
\iint_{\mathbb{R}^{N\times N}}\left(\frac{\lVert\mathbf{x}\rVert_2^2}{T^2}+T^2\lVert\mathbf{w}\rVert_2^2\right)\mathrm{C}f(\mathbf{x},\mathbf{w})\mathrm{d}\mathbf{x}\mathrm{d}\mathbf{w}\geq\frac{N}{2\pi}\lVert f\rVert_{L^2}^2
\end{equation}
for $T\in\mathbb{R}$, or equivalently
\begin{equation}\label{eq7.2}
\frac{\iint_{\mathbb{R}^{N\times N}}\left(\frac{\left\lVert\mathbf{x}-\mathbf{x}_f^0\right\rVert_2^2}{T^2}+T^2\left\lVert\mathbf{w}-\mathbf{w}_f^0\right\rVert_2^2\right)\mathrm{C}f(\mathbf{x},\mathbf{w})\mathrm{d}\mathbf{x}\mathrm{d}\mathbf{w}}{\lVert f\rVert_{L^2}^2}\geq\frac{N}{2\pi}
\end{equation}
for $T\in\mathbb{R}$. Here, the weak spreads in time-CCTFD and frequency-CCTFD domains are given by
\begin{equation}\label{eq7.3}
\frac{\iint_{\mathbb{R}^{N\times N}}\left\lVert\mathbf{x}-\mathbf{x}_f^0\right\rVert_2^2\mathrm{C}f(\mathbf{x},\mathbf{w})\mathrm{d}\mathbf{x}\mathrm{d}\mathbf{w}}{\lVert f\rVert_{L^2}^2}
\end{equation}
and
\begin{equation}\label{eq7.4}
\frac{\iint_{\mathbb{R}^{N\times N}}\left\lVert\mathbf{w}-\mathbf{w}_f^0\right\rVert_2^2\mathrm{C}f(\mathbf{x},\mathbf{w})\mathrm{d}\mathbf{x}\mathrm{d}\mathbf{w}}{\lVert f\rVert_{L^2}^2},
\end{equation}
respectively. These spreads are weak because there is the first power rather than the second power of the CCTFD found in them. Consequently, the weak Heisenberg's uncertainty principle of the CCTFD, shown in \eqref{eq7.2}, is none other than the classical one for the real-valued function, given by \eqref{eq5.2}.

Indeed, the standard definition of the spread is defined by the second power, such as $\left|f(\mathbf{x})\right|^2$ found in the time domain spread and $\left|\mathcal{F}f(\mathbf{w})\right|^2$ found in the frequency domain spread. In the current work, we focus on the UMITRSK-CCTFD, and strengthen \eqref{eq7.3} and \eqref{eq7.4} by replacing the first power $\mathrm{C}_{\mathrm{UMITRSK}}f(\mathbf{x},\mathbf{w})$ with the second power $\left|\mathrm{C}_{\mathrm{UMITRSK}}f(\mathbf{x},\mathbf{w})\right|^2$, giving birth to the spread $\Delta\mathbf{x}_{\mathrm{C}_{\mathrm{UMITRSK}},f}^2$ in the time-UMITRSK-CCTFD domain and the spread $\Delta\mathbf{w}_{\mathrm{C}_{\mathrm{UMITRSK}},f}^2$ in the frequency-UMITRSK-CCTFD domain. The standard Heisenberg's uncertainty principles of the UMITRSK-CCTFD derived are not the same as the classical ones. This is obvious, because the uncertainty product $\Delta\mathbf{x}_{\mathrm{C}_{\mathrm{UMITRSK}},f}^2\Delta\mathbf{w}_{\mathrm{C}_{\mathrm{UMITRSK}},f}^2$ in the UMITRSK-CCTFD domain is not a simple uncertainty product $\Delta\mathbf{x}_f^2\Delta\mathbf{w}_f^2$ in the FT domain, but a summation of two uncertainty products $\Delta\mathbf{x}_f^2\Delta\mathbf{w}_f^2,\Delta\mathbf{x}_{f\overline{\phi}}^2\Delta\mathbf{w}_{f\overline{\phi}}^2$ in the FT domain, as implied by \eqref{eq4.9}. To be specific, for the case of the real-valued function $f$ and the real-valued kernel $\phi=\pm1$, the derived uncertainty principle, shown in \eqref{eq6.1}, equals to the classical one for the real-valued function. But, as for the rest, the derived uncertainty principles, shown in \eqref{eq6.3}, \eqref{eq6.6} and \eqref{eq6.9}, differ essentially from the classical ones.

\section{Conclusion}\label{sec:8}

We have established the standard Heisenberg's uncertainty principles for classes of CCTFDs that share a common distribution property. Our results demonstrate that the unit modulus and $\mathbf{v}$-independent time translation, reversal and scaling invariant kernel constrains the time-frequency concentration and resolution of the CCTFD. The derived Heisenberg's inequalities on the uncertainty product in the UMITRSK-CCTFD domain are fourfold: the real-valued function $f$ and the real-valued kernel $\phi=\pm1$, the complex-valued function $f$ and the real-valued or complex-valued kernel $\phi\neq\pm\mathrm{e}^{2\pi\mathrm{i}\varphi_f}$, the real-valued function $f$ and the complex-valued kernel $\phi\neq\pm1$, and the complex-valued function $f$ and the complex-valued kernel $\phi=\pm\mathrm{e}^{2\pi\mathrm{i}\varphi_f}$. These results strengthen the celebrated Flandrin's result, giving rise to something new that differs essentially from the classical Heisenberg's uncertainty principles and indicating that a square integrable function cannot be sharply localized in both the time-UMITRSK-CCTFD domain and frequency-UMITRSK-CCTFD domain.

\section*{Appendix}

In this section, we prove all the theoretical results in the paper.

\subsection*{Appendix A. Proof of Lemma~\ref{Lem1}}

The result \eqref{eq2.1} of the one-dimensional case was given in \cite{Kor05} without proofs. Below, we provide the detailed proofs for the $N$-dimensional case. We rewrite the definition of the CCTFDs of $f,g$ in terms of integral form as
\begin{equation}\label{eqA.1}
\mathrm{C}f(\mathbf{x},\mathbf{w})=\iint_{\mathbb{R}^{N\times N}}f\left(\mathbf{z}+\frac{\mathbf{y}}{2}\right)\overline{f\left(\mathbf{z}-\frac{\mathbf{y}}{2}\right)}\left(\int_{\mathbb{R}^N}\phi(\mathbf{v},\mathbf{y})\mathrm{e}^{-2\pi\mathrm{i}\mathbf{v}(\mathbf{x}-\mathbf{z})^{\mathrm{T}}}\mathrm{d}\mathbf{v}\right)\mathrm{e}^{-2\pi\mathrm{i}\mathbf{y}\mathbf{w}^{\mathrm{T}}}\mathrm{d}\mathbf{z}\mathrm{d}\mathbf{y},
\tag{A.1}
\end{equation}
\begin{equation}\label{eqA.2}
\mathrm{C}g(\mathbf{x},\mathbf{w})=\iint_{\mathbb{R}^{N\times N}}g\left(\widehat{\mathbf{z}}+\frac{\widehat{\mathbf{y}}}{2}\right)\overline{g\left(\widehat{\mathbf{z}}-\frac{\widehat{\mathbf{y}}}{2}\right)}\left(\int_{\mathbb{R}^N}\phi(\widehat{\mathbf{v}},\widehat{\mathbf{y}})\mathrm{e}^{-2\pi\mathrm{i}\widehat{\mathbf{v}}(\mathbf{x}-\widehat{\mathbf{z}})^{\mathrm{T}}}\mathrm{d}\widehat{\mathbf{v}}\right)\mathrm{e}^{-2\pi\mathrm{i}\widehat{\mathbf{y}}\mathbf{w}^{\mathrm{T}}}\mathrm{d}\widehat{\mathbf{z}}\mathrm{d}\widehat{\mathbf{y}}.
\tag{A.2}
\end{equation}
Thanks to
\begin{equation}\label{eqA.3}
\int_{\mathbb{R}^N}\mathrm{e}^{-2\pi\mathrm{i}(\mathbf{y}-\widehat{\mathbf{y}})\mathbf{w}^{\mathrm{T}}}\mathrm{d}\mathbf{w}=\delta(\mathbf{y}-\widehat{\mathbf{y}}),
\tag{A.3}
\end{equation}
where $\delta(\cdot)$ denotes the Dirac delta operator for vectors, i.e., the product of element-wise Dirac deltas, and then by using the sifting property of Dirac delta functions, it follows that
\begin{align}\label{eqA.4}
\left\langle\mathrm{C}f,\mathrm{C}g\right\rangle_{(\mathbf{x},\mathbf{w})}=&\iiint_{\mathbb{R}^{N\times N\times N}}f\left(\mathbf{z}+\frac{\mathbf{y}}{2}\right)\overline{f\left(\mathbf{z}-\frac{\mathbf{y}}{2}\right)}\,\overline{g\left(\widehat{\mathbf{z}}+\frac{\mathbf{y}}{2}\right)}g\left(\widehat{\mathbf{z}}-\frac{\mathbf{y}}{2}\right)\notag\\
&\times\left(\iint_{\mathbb{R}^{N\times N}}\phi(\mathbf{v},\mathbf{y})\overline{\phi(\widehat{\mathbf{v}},\mathbf{y})}\mathrm{e}^{2\pi\mathrm{i}\mathbf{v}\mathbf{z}^{\mathrm{T}}}\mathrm{e}^{-2\pi\mathrm{i}\widehat{\mathbf{v}}\widehat{\mathbf{z}}^{\mathrm{T}}}\left(\int_{\mathbb{R}^N}\mathrm{e}^{-2\pi\mathrm{i}(\mathbf{v}-\widehat{\mathbf{v}})\mathbf{x}^{\mathrm{T}}}\mathrm{d}\mathbf{x}\right)\mathrm{d}\mathbf{v}\mathrm{d}\widehat{\mathbf{v}}\right)\mathrm{d}\mathbf{z}\mathrm{d}\widehat{\mathbf{z}}\mathrm{d}\mathbf{y}.
\tag{A.4}
\end{align}
Similarly, Eq.~\eqref{eqA.4} reduces to
\begin{equation}\label{eqA.5}
\left\langle\mathrm{C}f,\mathrm{C}g\right\rangle_{(\mathbf{x},\mathbf{w})}=\iint_{\mathbb{R}^{N\times N}}f\left(\mathbf{z}+\frac{\mathbf{y}}{2}\right)\overline{f\left(\mathbf{z}-\frac{\mathbf{y}}{2}\right)}\,\overline{g\left(\mathbf{z}+\frac{\mathbf{y}}{2}\right)}g\left(\mathbf{z}-\frac{\mathbf{y}}{2}\right)\mathrm{d}\mathbf{z}\mathrm{d}\mathbf{y},
\tag{A.5}
\end{equation}
with an assumption of $\left|\phi(\mathbf{v},\mathbf{y})\right|=1$. By taking the change of variables $(\mathbf{z},\mathbf{y})\rightarrow(\mathbf{z},\mathbf{y})\mathcal{P}^{-1}$, where $\mathcal{P}^{-1}=\begin{pmatrix}\frac{\mathbf{I}_N}{2}&\mathbf{I}_N\\\frac{\mathbf{I}_N}{2}&-\mathbf{I}_N\end{pmatrix}$, Eq.~\eqref{eqA.5} becomes
\begin{align}\label{eqA.6}
\left\langle\mathrm{C}f,\mathrm{C}g\right\rangle_{(\mathbf{x},\mathbf{w})}=&\iint_{\mathbb{R}^{N\times N}}f(\mathbf{z})\overline{g(\mathbf{z})}\,\overline{f(\mathbf{y})}g(\mathbf{y})\mathrm{d}\mathbf{z}\mathrm{d}\mathbf{y}\notag\\
=&\int_{\mathbb{R}^N}f(\mathbf{x})\overline{g(\mathbf{x})}\mathrm{d}\mathbf{x}\overline{\int_{\mathbb{R}^N}f(\mathbf{x})\overline{g(\mathbf{x})}\mathrm{d}\mathbf{x}}\notag\\
=&\left|\langle f,g\rangle_{\mathbf{x}}\right|^2.
\tag{A.6}
\end{align}
$\hfill\blacksquare$

\subsection*{Appendix B. Proof of Lemma~\ref{Lem2}}

We rewrite \eqref{eqA.1} as
\begin{equation}\label{eqB.1}
\mathrm{C}f(\mathbf{x},\mathbf{w})=\iint_{\mathbb{R}^{N\times N}}f\left(\mathbf{z}+\frac{\mathbf{y}}{2}\right)\overline{f\left(\mathbf{z}-\frac{\mathbf{y}}{2}\right)}\mathcal{F}_{\mathbf{v},1}\phi(\mathbf{x}-\mathbf{z},\mathbf{y})\mathrm{e}^{-2\pi\mathrm{i}\mathbf{y}\mathbf{w}^{\mathrm{T}}}\mathrm{d}\mathbf{z}\mathrm{d}\mathbf{y}.
\tag{B.1}
\end{equation}
By taking the change of variables $\mathbf{y}=2(\mathbf{z}-\mathbf{t})$, Eq.~\eqref{eqB.1} becomes
\begin{equation}\label{eqB.2}
\mathrm{C}f(\mathbf{x},\mathbf{w})=2^N\iint_{\mathbb{R}^{N\times N}}f(2\mathbf{z}-\mathbf{t})\overline{f(\mathbf{t})}\mathcal{F}_{\mathbf{v},1}\phi(\mathbf{x}-\mathbf{z},2(\mathbf{z}-\mathbf{t}))\mathrm{e}^{-4\pi\mathrm{i}(\mathbf{z}-\mathbf{t})\mathbf{w}^{\mathrm{T}}}\mathrm{d}\mathbf{z}\mathrm{d}\mathbf{t}.
\tag{B.2}
\end{equation}
By using the inverse formula of the FT, i.e.,
\begin{align}\label{eqB.3}
f(2\mathbf{z}-\mathbf{t})=&\mathcal{F}^{-1}\left(\mathcal{F}f\right)(2\mathbf{z}-\mathbf{t})\notag\\
=&\int_{\mathbb{R}^N}\mathcal{F}f(\mathbf{u})\mathrm{e}^{2\pi\mathrm{i}\mathbf{u}(2\mathbf{z}-\mathbf{t})^{\mathrm{T}}}\mathrm{d}\mathbf{u},
\tag{B.3}
\end{align}
we arrive the required result \eqref{eq3.1}.$\hfill\blacksquare$

\subsection*{Appendix C. Proof of Lemma~\ref{Lem3}}

$\phi(\mathbf{v},2(\mathbf{z}-\mathbf{t}))=\phi(\mathbf{t})$ implies that $\phi(\mathbf{v},\mathbf{y})=\phi(\mathbf{y})$, based on which $\mathcal{F}_{\mathbf{v},1}\phi(\mathbf{x}-\mathbf{z},\mathbf{y})$ found in \eqref{eq1.1} simplifies to $\phi(\mathbf{y})\delta(\mathbf{x}-\mathbf{z})$. Due to the sifting property of Dirac delta functions, the original time domain definition of the UMITRSK-CCTFD reduces to
\begin{equation}\label{eqC.1}
\mathrm{C}_{\mathrm{UMITRSK}}f(\mathbf{x},\mathbf{w})=\mathcal{F}_{\mathbf{y},2}\left(\mathfrak{T}_{\mathcal{P}}\left(f\otimes\overline{f}\right)(\mathbf{x},\mathbf{y})\phi(\mathbf{y})\right)(\mathbf{x},\mathbf{w}).
\tag{C.1}
\end{equation}
Substituting \eqref{eqC.1} into \eqref{eq4.4} yields
\begin{align}\label{eqC.2}
\mathbf{x}_{\mathrm{C}_{\mathrm{UMITRSK}},f}^0=&\frac{\left\langle\mathbf{x}\mathcal{F}_{\mathbf{y},2}\left(\mathfrak{T}_{\mathcal{P}}\left(f\otimes\overline{f}\right)\phi\right),\mathcal{F}_{\mathbf{y},2}\left(\mathfrak{T}_{\mathcal{P}}\left(f\otimes\overline{f}\right)\phi\right)\right\rangle_{(\mathbf{x},\mathbf{w})}}{\left\lVert f\right\rVert_{L^2}^4}\notag\\
=&\frac{\left\langle\mathbf{x}\mathfrak{T}_{\mathcal{P}}\left(f\otimes\overline{f}\right)\phi,\mathfrak{T}_{\mathcal{P}}\left(f\otimes\overline{f}\right)\phi\right\rangle_{(\mathbf{x},\mathbf{y})}}{\left\lVert f\right\rVert_{L^2}^4}.
\tag{C.2}
\end{align}
Thanks to $\left|\phi\right|=1$, Eq.~\eqref{eqC.2} becomes
\begin{equation}\label{eqC.3}
\mathbf{x}_{\mathrm{C}_{\mathrm{UMITRSK}},f}^0=\frac{\left\langle\mathbf{x}\mathfrak{T}_{\mathcal{P}}\left(f\otimes\overline{f}\right),\mathfrak{T}_{\mathcal{P}}\left(f\otimes\overline{f}\right)\right\rangle_{(\mathbf{x},\mathbf{y})}}{\left\lVert f\right\rVert_{L^2}^4}.
\tag{C.3}
\end{equation}
By taking the change of variables $(\mathbf{x},\mathbf{y})\rightarrow(\mathbf{x},\mathbf{y})\mathcal{P}^{-1}$, Eq.~\eqref{eqC.3} turns into
\begin{align}\label{eqC.4}
\mathbf{x}_{\mathrm{C}_{\mathrm{UMITRSK}},f}^0=&\frac{\left\langle(\mathbf{x}+\mathbf{y})f\otimes\overline{f},f\otimes\overline{f}\right\rangle_{(\mathbf{x},\mathbf{y})}}{2\left\lVert f\right\rVert_{L^2}^4}\notag\\
=&\frac{\left\langle\mathbf{x}f\otimes\overline{f},f\otimes\overline{f}\right\rangle_{(\mathbf{x},\mathbf{y})}}{2\left\lVert f\right\rVert_{L^2}^4}+\frac{\left\langle f\otimes\mathbf{y}\overline{f},f\otimes\overline{f}\right\rangle_{(\mathbf{x},\mathbf{y})}}{2\left\lVert f\right\rVert_{L^2}^4}\notag\\
=&\frac{\left\langle\mathbf{x}f,f\right\rangle_{\mathbf{x}}\left\lVert\overline{f}\right\rVert_{L^2}^2}{2\left\lVert f\right\rVert_{L^2}^4}+\frac{\left\langle\mathbf{y}\overline{f},\overline{f}\right\rangle_{\mathbf{y}}\left\lVert f\right\rVert_{L^2}^2}{2\left\lVert f\right\rVert_{L^2}^4}\notag\\
=&\mathbf{x}_f^0,
\tag{C.4}
\end{align}
which indicates that the moment vector $\mathbf{x}_{\mathrm{C}_{\mathrm{UMITRSK}},f}^0$ in the time-UMITRSK-CCTFD domain is none other than the moment vector $\mathbf{x}_f^0$ in the time domain. Then, the spread $\Delta\mathbf{x}_{\mathrm{C}_{\mathrm{UMITRSK}},f}^2$ in the time-UMITRSK-CCTFD domain becomes
\begin{align}\label{eqC.5}
\Delta\mathbf{x}_{\mathrm{C}_{\mathrm{UMITRSK}},f}^2=&\frac{\left\lVert\left(\mathbf{x}-\mathbf{x}_f^0\right)\mathrm{C}_{\mathrm{UMITRSK}}f\right\rVert_{L^2}^2}{\left\lVert f\right\rVert_{L^2}^4}\notag\\
=&\frac{\left\lVert\left(\mathbf{x}-\mathbf{x}_f^0\right)\mathcal{F}_{\mathbf{y},2}\left(\mathfrak{T}_{\mathcal{P}}\left(f\otimes\overline{f}\right)\phi\right)\right\rVert_{L^2}^2}{\left\lVert f\right\rVert_{L^2}^4}\notag\\
=&\frac{\left\lVert\left(\mathbf{x}-\mathbf{x}_f^0\right)\mathfrak{T}_{\mathcal{P}}\left(f\otimes\overline{f}\right)\phi\right\rVert_{L^2}^2}{\left\lVert f\right\rVert_{L^2}^4}\notag\\
=&\frac{\left\lVert\left(\mathbf{x}-\mathbf{x}_f^0\right)\mathfrak{T}_{\mathcal{P}}\left(f\otimes\overline{f}\right)\right\rVert_{L^2}^2}{\left\lVert f\right\rVert_{L^2}^4}\notag\\
=&\frac{\left\lVert\left(\frac{\mathbf{x}+\mathbf{y}}{2}-\mathbf{x}_f^0\right)f\otimes\overline{f}\right\rVert_{L^2}^2}{\left\lVert f\right\rVert_{L^2}^4}\notag\\
=&\frac{\left\lVert\left(\mathbf{x}-\mathbf{x}_f^0\right)f\otimes\overline{f}+f\otimes\left(\mathbf{y}-\mathbf{x}_f^0\right)\overline{f}\right\rVert_{L^2}^2}{4\left\lVert f\right\rVert_{L^2}^4}\notag\\
=&\frac{\left\lVert\left(\mathbf{x}-\mathbf{x}_f^0\right)f\otimes\overline{f}\right\rVert_{L^2}^2}{4\left\lVert f\right\rVert_{L^2}^4}+\frac{\left\lVert f\otimes\left(\mathbf{y}-\mathbf{x}_f^0\right)\overline{f}\right\rVert_{L^2}^2}{4\left\lVert f\right\rVert_{L^2}^4}\notag\\
=&\frac{\left\lVert\left(\mathbf{x}-\mathbf{x}_f^0\right)f\right\rVert_{L^2}^2\left\lVert\overline{f}\right\rVert_{L^2}^2}{4\left\lVert f\right\rVert_{L^2}^4}+\frac{\left\lVert\left(\mathbf{y}-\mathbf{x}_f^0\right)\overline{f}\right\rVert_{L^2}^2\left\lVert f\right\rVert_{L^2}^2}{4\left\lVert f\right\rVert_{L^2}^4}\notag\\
=&\frac{\Delta\mathbf{x}_f^2}{2}.
\tag{C.5}
\end{align}
$\hfill\blacksquare$

\subsection*{Appendix D. Proof of Lemma~\ref{Lem4}}

Substituting \eqref{eq3.2} into \eqref{eq4.6} yields
\begin{align}\label{eqD.1}
\mathbf{w}_{\mathrm{C}_{\mathrm{UMITRSK}},f}^0=&\frac{2^N\left\langle\mathbf{w}\mathcal{F}_{\mathbf{u},2}\mathfrak{T}_{\mathcal{Q}}\left(\mathcal{F}f\otimes\overline{\mathcal{F}\left(f\overline{\phi}\right)}\right)(\mathbf{w},-2\mathbf{x}),\mathcal{F}_{\mathbf{u},2}\mathfrak{T}_{\mathcal{Q}}\left(\mathcal{F}f\otimes\overline{\mathcal{F}\left(f\overline{\phi}\right)}\right)(\mathbf{w},-2\mathbf{x})\right\rangle_{(\mathbf{x},\mathbf{w})}}{\left\lVert f\right\rVert_{L^2}^4}\notag\\
=&\frac{\left\langle\mathbf{w}\mathfrak{T}_{\mathcal{Q}}\left(\mathcal{F}f\otimes\overline{\mathcal{F}\left(f\overline{\phi}\right)}\right),\mathfrak{T}_{\mathcal{Q}}\left(\mathcal{F}f\otimes\overline{\mathcal{F}\left(f\overline{\phi}\right)}\right)\right\rangle_{(\mathbf{w},\mathbf{u})}}{\left\lVert f\right\rVert_{L^2}^4}.
\tag{D.1}
\end{align}
By taking the change of variables $(\mathbf{w},\mathbf{u})\rightarrow(\mathbf{w},\mathbf{u})\mathcal{Q}^{-1}$, where $\mathcal{Q}^{-1}=\begin{pmatrix}\frac{\mathbf{I}_N}{2}&\mathbf{I}_N\\\frac{\mathbf{I}_N}{2}&\mathbf{0}_N\end{pmatrix}$, Eq.~\eqref{eqD.1} simplifies to
\begin{align}\label{eqD.2}
\mathbf{w}_{\mathrm{C}_{\mathrm{UMITRSK}},f}^0=&\frac{\left\langle(\mathbf{w}+\mathbf{u})\mathcal{F}f\otimes\overline{\mathcal{F}\left(f\overline{\phi}\right)},\mathcal{F}f\otimes\overline{\mathcal{F}\left(f\overline{\phi}\right)}\right\rangle_{(\mathbf{w},\mathbf{u})}}{2\left\lVert f\right\rVert_{L^2}^4}\notag\\
=&\frac{\left\langle\mathbf{w}\mathcal{F}f\otimes\overline{\mathcal{F}\left(f\overline{\phi}\right)},\mathcal{F}f\otimes\overline{\mathcal{F}\left(f\overline{\phi}\right)}\right\rangle_{(\mathbf{w},\mathbf{u})}}{2\left\lVert f\right\rVert_{L^2}^4}+\frac{\left\langle\mathcal{F}f\otimes\mathbf{u}\overline{\mathcal{F}\left(f\overline{\phi}\right)},\mathcal{F}f\otimes\overline{\mathcal{F}\left(f\overline{\phi}\right)}\right\rangle_{(\mathbf{w},\mathbf{u})}}{2\left\lVert f\right\rVert_{L^2}^4}\notag\\
=&\frac{\left\langle\mathbf{w}\mathcal{F}f,\mathcal{F}f\right\rangle_{\mathbf{w}}\left\lVert\overline{\mathcal{F}\left(f\overline{\phi}\right)}\right\rVert_{L^2}^2}{2\left\lVert f\right\rVert_{L^2}^4}+\frac{\left\langle\mathbf{u}\overline{\mathcal{F}\left(f\overline{\phi}\right)},\overline{\mathcal{F}\left(f\overline{\phi}\right)}\right\rangle_{\mathbf{u}}\left\lVert f\right\rVert_{L^2}^2}{2\left\lVert f\right\rVert_{L^2}^4}.
\tag{D.2}
\end{align}
Because of Parseval's relation of the FT and $\left|\phi\right|=1$, it follows that $\left\lVert\overline{\mathcal{F}\left(f\overline{\phi}\right)}\right\rVert_{L^2}=\left\lVert\mathcal{F}\left(f\overline{\phi}\right)\right\rVert_{L^2}=\left\lVert f\overline{\phi}\right\rVert_{L^2}=\left\lVert f\right\rVert_{L^2}$. Then, Eq.~\eqref{eqD.2} becomes
\begin{align}\label{eqD.3}
\mathbf{w}_{\mathrm{C}_{\mathrm{UMITRSK}},f}^0=&\frac{\left\langle\mathbf{w}\mathcal{F}f,\mathcal{F}f\right\rangle_{\mathbf{w}}}{2\left\lVert f\right\rVert_{L^2}^2}+\frac{\left\langle\mathbf{u}\mathcal{F}\left(f\overline{\phi}\right),\mathcal{F}\left(f\overline{\phi}\right)\right\rangle_{\mathbf{u}}}{2\left\lVert f\overline{\phi}\right\rVert_{L^2}^2}\notag\\
=&\frac{\mathbf{w}_f^0+\mathbf{w}_{f\overline{\phi}}^0}{2},
\tag{D.3}
\end{align}
which indicates that the moment vector $\mathbf{w}_{\mathrm{C}_{\mathrm{UMITRSK}},f}^0$ in the frequency-UMITRSK-CCTFD domain is none other than a summation of the moment vectors $\mathbf{w}_f^0,\mathbf{w}_{f\overline{\phi}}^0$ in the frequency domain, regardless of a multiplier $\frac{1}{2}$. Then, the spread $\Delta\mathbf{w}_{\mathrm{C}_{\mathrm{UMITRSK}},f}^2$ in the frequency-UMITRSK-CCTFD domain turns into
\begin{align}\label{eqD.4}
\Delta\mathbf{w}_{\mathrm{C}_{\mathrm{UMITRSK}},f}^2=&\frac{\left\lVert\left(\mathbf{w}-\frac{\mathbf{w}_f^0+\mathbf{w}_{f\overline{\phi}}^0}{2}\right)\mathrm{C}_{\mathrm{UMITRSK}}f\right\rVert_{L^2}^2}{\left\lVert f\right\rVert_{L^2}^4}\notag\\
=&\frac{\left\lVert\left(\mathbf{w}-\frac{\mathbf{w}_f^0+\mathbf{w}_{f\overline{\phi}}^0}{2}\right)\mathfrak{T}_{\mathcal{Q}}\left(\mathcal{F}f\otimes\overline{\mathcal{F}\left(f\overline{\phi}\right)}\right)\right\rVert_{L^2}^2}{\left\lVert f\right\rVert_{L^2}^4}\notag\\
=&\frac{\left\lVert\left(\mathbf{w}-\mathbf{w}_f^0\right)\mathcal{F}f\otimes\overline{\mathcal{F}\left(f\overline{\phi}\right)}+\mathcal{F}f\otimes\left(\mathbf{u}-\mathbf{w}_{f\overline{\phi}}^0\right)\overline{\mathcal{F}\left(f\overline{\phi}\right)}\right\rVert_{L^2}^2}{4\left\lVert f\right\rVert_{L^2}^4}\notag\\
=&\frac{\left\lVert\left(\mathbf{w}-\mathbf{w}_f^0\right)\mathcal{F}f\right\rVert_{L^2}^2}{4\left\lVert f\right\rVert_{L^2}^2}+\frac{\left\lVert\left(\mathbf{u}-\mathbf{w}_{f\overline{\phi}}^0\right)\mathcal{F}\left(f\overline{\phi}\right)\right\rVert_{L^2}^2}{4\left\lVert f\overline{\phi}\right\rVert_{L^2}^2}\notag\\
=&\frac{\Delta\mathbf{w}_f^2+\Delta\mathbf{w}_{f\overline{\phi}}^2}{4}.
\tag{D.4}
\end{align}
$\hfill\blacksquare$

\section*{Conflict of interest statement}

The author declares that there is no conflict of interests to this work.

\bibliography{mybib}{}
\bibliographystyle{plain}
\end{document}